\documentclass[amsmath,amssymb,aps,prd,twocolumn,superscriptaddress,10pt]{revtex4-2}
\usepackage{graphicx}
\usepackage{dcolumn}
\usepackage{subfigure}
\usepackage{color}
\usepackage{footnote}
\usepackage{url}
\usepackage{rotating}
\usepackage{xspace}

\usepackage[dvipsnames]{xcolor}
\usepackage[colorlinks,urlcolor=Blue,citecolor=Blue,linkcolor=Blue,pdfusetitle]{hyperref}

\newcommand\apjl{The Astrophysical Journal Letters}
\newcommand\aj{The Astronomical Journal}
\newcommand\mnras{Monthly Notices of the Royal Astronomical Society}
\newcommand\aap{Astronomy and Astrophysics}

\begin{document}

\title{Effects of spin-orbit coupling on gravitational waveforms from a triaxial non-aligned neutron star in a binary system}

\author{Wen-Fan Feng}
\affiliation{MOE Key Laboratory of Fundamental Physical Quantities Measurements, Hubei Key Laboratory of Gravitation and Quantum Physics, PGMF, Department of Astronomy and School of Physics, Huazhong University of Science and Technology, Wuhan 430074, China}

\author{Tan Liu}
\email{lewton@mail.ustc.edu.cn}
\affiliation{MOE Key Laboratory of Fundamental Physical Quantities Measurements, Hubei Key Laboratory of Gravitation and Quantum Physics, PGMF, Department of Astronomy and School of Physics, Huazhong University of Science and Technology, Wuhan 430074, China}
\affiliation{School of Physics, Hubei University, Wuhan 430062, China}

\author{Jie-Wen Chen}
\affiliation{MOE Key Laboratory of Fundamental Physical Quantities Measurements, Hubei Key Laboratory of Gravitation and Quantum Physics, PGMF, Department of Astronomy and School of Physics, Huazhong University of Science and Technology, Wuhan 430074, China}
\affiliation{National Time Service Center, Chinese Academy of Sciences, Xi’an 710600, China}
\affiliation{Key Laboratory of Time and Frequency Primary Standards, Chinese Academy of Sciences, Xi’an 710600, China}

\author{Yan Wang}
\email{ywang12@hust.edu.cn}
\affiliation{MOE Key Laboratory of Fundamental Physical Quantities Measurements, Hubei Key Laboratory of Gravitation and Quantum Physics, PGMF, Department of Astronomy and School of Physics, Huazhong University of Science and Technology, Wuhan 430074, China}

\author{Soumya D.~Mohanty}
\affiliation{Department of Physics and Astronomy, University of Texas Rio Grande Valley, 
Brownsville, Texas 78520, USA}
\affiliation{Department of Physics, IIT Hyderabad, Kandai, Telangana-502284, India}

\date{\today}

\begin{abstract}

Spinning neutron stars (NSs) can emit continuous gravitational waves (GWs) that carry a wealth of information about the compact object. If such a signal is detected, it will provide us with new insight into the physical properties of matter under extreme conditions. 
According to binary population synthesis simulations, future space-based GW detectors, such as LISA and TianQin, can potentially detect some double NSs in tight binaries with orbital periods shorter than 10 minutes. 
The possibility of a successful directed search for continuous GWs from the spinning NS in such a binary system identified by LISA/TianQin will be significantly increased with the proposed next-generation ground-based GW observatories, such as Cosmic Explorer and Einstein Telescope. 
Searching for continuous GWs from such a tight binary system requires highly accurate waveform templates that account for the interaction of the NS with its companion.
In this spirit, we derive analytic approximations that describe the GWs emitted by a triaxial non-aligned NS in a binary system in which the effects of spin-orbit coupling have been incorporated.
The difference with the widely used waveform for the isolated NS is estimated and the parameter estimation accuracy of an example signal using Cosmic Explorer is calculated. 
For a typical tight double NS system with a 6~min orbital period, the angular frequency correction of the spinning NS in this binary due to spin precession is $\sim 10^{-6}~{\rm Hz}$, which is in the same order of magnitude as the angular frequency of orbital precession. 
The fitting factor between the waveforms with and without spin precession will drop to less than 0.97 after a few days ($\sim 10^5~{\rm s}$). 
We find that spin-orbit coupling has the potential to improve the accuracy of parameter estimation, especially for the binary inclination angle and spin precession cone opening angle, by up to 3 orders of magnitude.

\end{abstract}

\maketitle

\section{Introduction}

Rapidly spinning neutron stars (NSs) are promising sources of a long-lasting form of gravitational waves (GWs), namely continuous waves (CWs) \citep{Sieniawska2019, Riles2017, Lasky2015, 2022arXiv220606447R, Wette2023}. Detecting these potential CWs would help us solve some of the mysteries in NS physics, such as NS's equation of state, deformability, and magnetic field \cite{Pitkin2011, Soldateschi2021, Lu2022}.

There are two types of simplified waveforms that are commonly used in current searches for CWs emitted by NSs (modeled as Newtonian rigid bodies) with Advanced LIGO \cite{advancedLIGO2015} and Advanced Virgo \cite{AdvancedVirgo2015}. One is the mass quadrupole mode with a frequency at twice the  rotation frequency of the pulsar, which comes from a triaxial rigid body rotating about one of its principal axes with assumed principal moments of inertia $I_1<I_2<I_3$ (hereafter referred to as the  triaxial aligned waveforms, e.g., \cite{LIGO2022isolatedO3, LIGOisolatedNS2021, Abbott2022Narrowband, Abbott2022CasA}); 
the other is the mode with frequencies at both once and twice the rotation frequency, which comes from an axisymmetric freely precessing rigid body with assumed $I_1=I_2 \ne I_3$ (hereafter referred to as the biaxial waveforms, e.g., \cite{LIGOYoung2021ApJ, Sun2019, Jaranowski1998}). 
Similar two-frequency mode searches are also performed in \cite{Abbott2022TwoHarmonics, LIGO2019TwoHarmonics, Pitkin2015MNRAS, Jones2015MNRAS, Jones2010MNRAS, Galtsov1984, Bonazzola1996}. The GW emission due to $r$-modes \cite{Lindblom1998, Andersson1998ApJ} in a rotating perfect fluid star is not the subject of our work.
So far, no credible detection has been reported in these searches~\citep{Wette2023}.

The more general waveforms that come from a freely precessing triaxial rigid body (hereafter referred to as the triaxial non-aligned waveforms) were first calculated by Zimmermann \cite{Zimmermann1980}.
The dominant waveform components are obtained by expanding the quadrupole moment formula in terms of small parameters, such as wobble angle, oblateness, and non-axisymmetry parameters. 
These waveforms have been extended to include higher than first-order expansion terms of the wobble angle and non-axisymmetry \cite{Broeck2005, Gao2020} in order to extract more physical information. 

In addition to the waveform modeling of isolated NS discussed above, there are also considerations about the NS located in a binary system since the electromagnetic observations show that nearly half of the known pulsars within the most sensitive band of the ground-based GW detectors belong to binary systems \cite{Covas2019, ATNF_website, Manchester2005}. 
Some search schemes are proposed for this type of CWs~\cite{LIGONSinBinary2021, Covas2020, Covas2019, Zhang2021, Leaci2015}. However, the waveform model used in these searches is obtained by simply incorporating the Doppler frequency modulation into the phase of the triaxial aligned waveforms emitted by the isolated NS.

For future space-borne GW detectors, the detectability and parameter estimation accuracy of double NS systems that will merge within the next 10 Myr have been studied for LISA \cite{Andrews2020} and TianQin \cite{Feng2023}. Based on the merger rate density ($920 ~\mathrm{Gpc^{-3}~yr^{-1}}$) inferred from GWTC-1 \cite{GWTC12019}, about 300 double NS systems are expected to be detected in the mHz band during the 4-year observation period, including binaries with orbital periods shorter than 10 minutes. 
Proposed next-generation ground-based GW observatories, such as Cosmic Explorer \cite{CosmicExplorer2022} and Einstein Telescope \cite{ET2010}, are expected to operate concurrently with LISA and TianQin in the 2030s.
Searching for CWs from the spinning NS in such a tight system identified by LISA and/or TianQin requires consideration of the interaction of the rapidly spinning NS with its companion. 

In this paper, we incorporate the effects of spin-orbit coupling to the GWs emitted by the spinning NS in a circular orbital binary and extend the triaxial aligned NS to a general triaxial non-aligned NS. 
Other effects, such as magnetic dipole field \cite{Davis1970}, tidal interaction \cite{Bildsten1992}, and radiation reaction \cite{Apostolatos1994} are neglected in the current work. 
Spin-orbit coupling causes spin precession and orbital precession around the total angular momentum \cite{Apostolatos1994}.
Similar to the treatment in previous works, such as that of LIGO \cite{LIGONSinBinary2021}, the GWs emitted by the spinning NS in a binary are obtained by simply incorporating the Doppler frequency modulation (including the effects of orbital precession) into the phase of the triaxial non-aligned waveforms of the NS with spin precession. 
In contrast to the isolated case (neglecting the electromagnetic and the gravitational radiation-reaction torque as in \cite{Zimmermann1980}), the spin angular momentum of the NS is no longer constant in the binary. Instead, it will be precessed due to the spin-orbit coupling. 
We analytically solve the spin precession equation for the NS using the perturbation method to obtain the spin angular frequency evolution and calculate the waveforms based on the quadrupole moment formula. Next, the waveforms are expanded into some simple components in the small parameter case for the subsequent analysis of CW detection. Finally, using these easy-to-use waveform components, we investigate the impact of spin-orbit coupling on the parameter estimation accuracy of the spinning NS.
Calculations along these lines yield the following results: (i) The waveforms of the NS undergoing spin precession will deviate from the isolated ones after a few days ($\sim 10^5~{\rm s}$) when the fitting factor between the two waveforms drops to less than 0.97.
(ii) Spin-orbit coupling has the potential to improve the parameter estimation accuracy, specifically for the cosine of the binary inclination $\cos\iota$ and spin precession cone opening angle $\theta_S$, by up to 3 orders of magnitude. 

The rest of this paper is organized as follows. In  Sec.~\ref{sec:isolatedNSwaveform}, we briefly review the mathematical formalism for GWs from an isolated spinning NS, which will be used for subsequent calculations for NS in a binary system.
Analytical approximations for the GWs from a spinning NS in a binary system, taking into account spin-orbit coupling effects, are given in Sec.~\ref{sec:NSinbinarywaveform}. 
The comparison of results derived using waveforms 
with and without spin precession is given in Sec.~\ref{sec:comparisonwithisolated}.
The parameter estimation accuracy of the waveforms with and without spin-orbit coupling using Cosmic Explorer \cite{CosmicExplorer2022} are given in Sec.~\ref{sec:detection}. 
Our conclusions are discussed in Sec.~\ref{sec:conclusion}. 
Some details of our calculation have been relegated to the appendix in order to keep the main ideas of the paper as clear as possible.

\section{Gravitational waveforms from isolated NS}
\label{sec:isolatedNSwaveform} 

\begin{figure}[!htbp]
\centering
\includegraphics[scale=0.55]{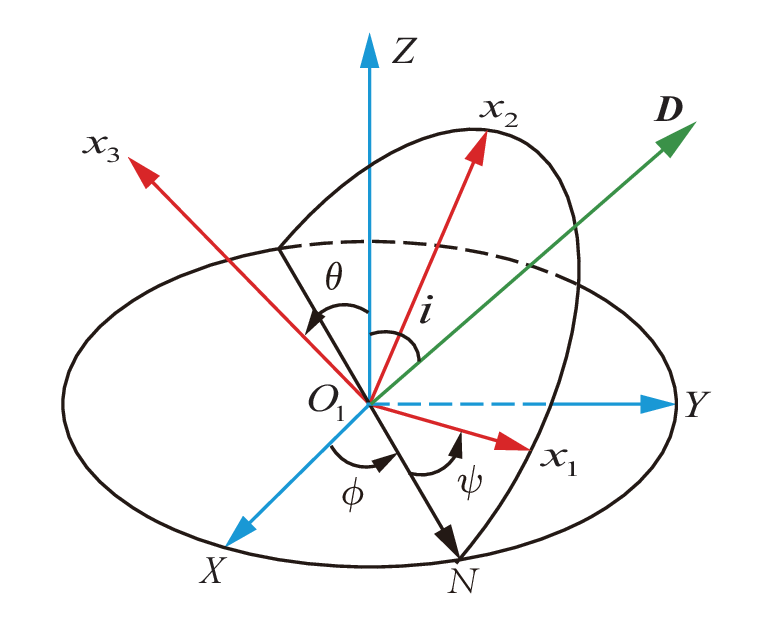}
\caption{The inertial coordinate system $(X, Y, Z)$ and the body coordinate system $(x_1,x_2,x_3)$ are both centered on the center of mass ($O_1$) of the isolated  NS.
The $x_1-x_2$ plane intersects the $X-Y$ plane at the line of nodes $O_1 N$.
($\theta,\phi,\psi$) are Euler angles. The distant observer with position vector $\boldsymbol{D}$ is assumed in the $Y-Z$ plane at colatitude $i$ from the $Z$ axis.}
\label{fig:IsolatedNS}
\end{figure}

Since the waveforms emitted by spinning NS undergoing spin-orbit coupling are based on the waveforms emitted by the isolated NS, we will first discuss the case for the isolated NS. 
Following the conventions of Landau and Lifshitz \cite{LANDAU1976} and Zimmermann \cite{Zimmermann1980}, in Fig.~\ref{fig:IsolatedNS}, the inertial coordinate system is denoted as $(X, Y, Z)$ with basis vectors $(\boldsymbol{e}_x,\boldsymbol{e}_y,\boldsymbol{e}_z)$ and $\boldsymbol{e}_z$ along the body's angular momentum, and the body coordinate system $(x_1,x_2,x_3)$ with basis vectors $(\boldsymbol{e}_1,\boldsymbol{e}_2,\boldsymbol{e}_3)$ parallel to the eigenvectors of the body's moment of inertia tensor and satisfying $I_3>I_2 \geq I_1$. The origins of the two systems are placed at the center of mass of the NS.
The Euler angles ($\theta,\phi,\psi$) describe the orientation of the body coordinate system with respect to the inertial coordinate system. 
We use the Latin subscripts (e.g., $x,y,z$) for components evaluated in the inertial coordinate system, and the Greek ones (e.g., $\mu,\nu$) in the body coordinate system. 

The metric perturbation under the transverse-traceless gauge can be written in terms of two GW polarizations, $h_{j k}^{\mathrm{TT}}=h_{+}\left(\hat{e}_{+}\right)_{j k}+h_{\times}\left(\hat{e}_{\times}\right)_{j k}$, with the polarization tensors defined as 
\begin{equation}
\label{eq_polatensor}
\hat{e}_{+} \equiv \hat{v} \otimes \hat{v}-\hat{w} \otimes \hat{w}  \,, \quad
\hat{e}_{\times} \equiv \hat{v} \otimes \hat{w}+\hat{w} \otimes \hat{v} \,,
\end{equation}
where $\hat{v}$ and $\hat{w}$ are the transverse basis vectors perpendicular to the wave's propagation direction, and $\otimes$ denotes the tensor product.
Without loss of generality, we assume that the observer is located in the $Y-Z$ plane with colatitude $i$ from the $Z$ axis and distance $D=|\boldsymbol{D}|$. 
In this configuration, 
\begin{equation}\label{eq_transbasis}
\hat{v} \equiv \hat{e}_{y} \cos i-\hat{e}_{z} \sin i \,,\quad  \hat{w} \equiv -\hat{e}_{x} \,, 
\end{equation}
and the two GW polarizations can be written as  \cite{Zimmermann1980}
\begin{subequations}
\label{eq:waveform_h}
\begin{align}\label{eq:waveform_hplusRA}
h_{+} &= -\frac{G}{c^4 D} [(R_{y\mu} \cos i - R_{z\mu} \sin i) \\ \nonumber
&\times (R_{y\nu} \cos i - R_{z\nu} \sin i)-R_{x\mu} R_{x\nu}] A_{\mu \nu} \,, \\  \label{eq:waveform_hcrossRA}
h_{\times} &= \frac{2 G}{c^4 D} (R_{y\mu} \cos i - R_{z\mu} \sin i) R_{x\nu} A_{\mu \nu}  \,,
\end{align}
\end{subequations}
where Einstein summation is performed for $\mu$ and $\nu$, both of which take the values $\{1,2,3\}$. The components of the symmetric matrix $A_{\mu\nu}$ are given by \cite{Zimmermann1980}
\begin{subequations}
\label{eq:Amatrix}
\begin{align}\label{eq:A11}
A_{11} &= 2\left(\Delta_{2} \Omega_{2}^{2}-\Delta_{3} \Omega_{3}^{2}\right) \,,\\ 
A_{22} &= 2\left(\Delta_{3} \Omega_{3}^{2}-\Delta_{1} \Omega_{1}^{2}\right) \,,\\ 
A_{33} &= 2\left(\Delta_{1} \Omega_{1}^{2}-\Delta_{2} \Omega_{2}^{2}\right) \,,\\ 
A_{12} &= A_{21} = \left(\Delta_{1}-\Delta_{2}\right) \Omega_{1} \Omega_{2}+\Delta_{3} \dot{\Omega}_{3}  \,,\\
A_{23} &= A_{32} = \left(\Delta_{2}-\Delta_{3}\right) \Omega_{2} \Omega_{3}+\Delta_{1} \dot{\Omega}_{1}   \,,\\ \label{eq:A31}
A_{31} &= A_{13} = \left(\Delta_{3}-\Delta_{1}\right) \Omega_{3} \Omega_{1}+\Delta_{2} \dot{\Omega}_{2}   \,,
\end{align}
\end{subequations}
with $\Delta_{1} \equiv I_{2}-I_{3}$, $\Delta_{2} \equiv I_{3}-I_{1}$, $\Delta_{3} \equiv I_{1}-I_{2}$. 
Here, $(\Omega_1,\Omega_2,\Omega_3)$ denotes the angular frequency of NS in the body coordinate system. 
The rotation matrix that transforms from the body coordinate system to the inertial coordinate system (e.g., $R_{x2}$ denotes the entry in row $x=1$ and column $\mu=2$) is 
\begin{widetext}
\begin{align}\label{Rmatrix}
R= \left(
\begin{array}{ccc}
 \cos \psi \cos \phi-\cos \theta \sin \psi \sin \phi    & -\cos \theta \cos \psi \sin \phi-\sin \psi \cos \phi    & \sin \theta \sin \phi  \\
 \cos \theta \sin \psi \cos \phi +\cos \psi \sin \phi   & \cos \theta \cos \psi  \cos \phi -\sin \psi \sin \phi   & -\sin \theta \cos \phi \\
 \sin \theta  \sin \psi                                 & \sin \theta \cos \psi                                   & \cos \theta             \\
\end{array}
\right) \,.
\end{align}
\end{widetext}

According to Euler’s equations of free rotation of a rigid body and the initial conditions ($\Omega_1(0) = a, \Omega_2(0) =0, \Omega_3(0) = b$), the angular frequencies in the body coordinate system are \cite{Zimmermann1980}
\begin{subequations}\label{eq_iso_Omega}
\begin{align}
\Omega_{1} &= a \operatorname{cn}(\tau, m)  \,,\\ 
\Omega_{2} &= a \left[\frac{I_{1}\left(I_{3}-I_{1}\right)}{I_{2}\left(I_{3}-I_{2}\right)}\right]^{1 / 2} \operatorname{sn}(\tau, m) \,,\\ 
\Omega_{3} &= b \operatorname{dn}(\tau, m)  \,,
\end{align}
\end{subequations}
where $\rm cn$, $\rm sn$ and $\rm dn$ are Jacobian elliptic functions \cite{Functions1970} with the parameters  
\begin{align}\label{eq_tau}
\tau &= b t\left[\frac{\left(I_{3}-I_{2}\right)\left(I_{3}-I_{1}\right)}{I_{1} I_{2}}\right]^{1 / 2} ,\\ \label{eq_m}
 m &= \frac{\left(I_{2}-I_{1}\right) I_{1} a^{2}}{\left(I_{3}-I_{2}\right) I_{3} b^{2}} \,.
\end{align}
The Euler angles can be expressed in terms of Jacobian elliptic functions and the fourth theta functions \cite{Functions1970} ($\vartheta_{4}$ and their derivatives $\vartheta_{4}^{\prime}$):
\begin{subequations}\label{eq_Euler_iso}
\begin{align}
\cos \theta &= \frac{I_{3} b}{S} \operatorname{dn} (\tau, m) \,,\\ 
\tan \psi &= \left[\frac{I_{1}\left(I_{3}-I_{2}\right)}{I_{2}\left(I_{3}-I_{1}\right)}\right]^{1 / 2} \frac{\operatorname{cn}(\tau, m)}{\operatorname{sn}(\tau, m)} \,, \\
\phi &= \phi_{1}+\phi_{2} \,, 
\end{align}
\end{subequations}
with $S$ being the magnitude of the spin angular momentum of the NS and $\phi_{1,2}$ given by \cite{Gao2020}
\begin{subequations}\label{eq:phi12}
\begin{align}
&\exp \left[2 \mathrm{i} \phi_{1}(t)\right]=\frac{\vartheta_{4}\left(\frac{2 \pi t}{T}+\mathrm{i} \pi \alpha, q \right)}{\vartheta_{4}\left(\frac{2 \pi t}{T}-\mathrm{i} \pi \alpha , q \right)} \,,\\ 
&\phi_{2}=\frac{2 \pi t}{T^{\prime}}=\left[\frac{S}{I_{1}}+\frac{2 \pi \mathrm{i}}{T} \frac{\vartheta_{4}^{\prime}(\mathrm{i} \pi \alpha, q)}{\vartheta_{4}(\mathrm{i} \pi \alpha, q)} \right]t  \,.
\end{align}
\end{subequations}
Here, $T$ is the period of the angular frequency in the body coordinate system, 
\begin{equation}\label{eq_T}
T =\frac{4 K(m)}{b}\left[\frac{I_{1} I_{2}}{\left(I_{3}-I_{2}\right)\left(I_{3}-I_{1}\right)}\right]^{1 / 2}  \,. 
\end{equation}
$\alpha$ satisfies ${\rm sn}[2 \mathrm{i} \alpha K(m)] =\mathrm{i} I_{3} b/(I_{1} a)$. 
$\mathrm{i}$ is the imaginary unit. 
$q =\exp [-\pi K(1-m)/K(m)]$, where $K(m)$ is the complete elliptic integral of the first kind \cite{Functions1970}. 
Since $\cos\phi_2$ has a period $T^{\prime}$ which is generally not commensurate with $T$, the motion of the NS is usually nonperiodic. When the NS becomes axisymmetric, $T^{\prime} \to 2\pi I_1/S$ \cite{Zimmermann1980}.

Explicit waveforms are usually expressed in a series expansion of some small parameters \cite{Zimmermann1980, Broeck2005, Gao2020}.
To facilitate the following calculation, we define the spinning NS's free precession angular frequency and rotation angular frequency
\begin{align}\label{eq_preandrotfreq}
\Omega_{\rm p} \equiv \frac{2\pi}{T} \,,\quad
\Omega_{\rm r} \equiv \frac{2\pi}{T^{\prime}}-\frac{2\pi}{T} \,,
\end{align}
and three parameters that characterize NS's properties  
\begin{align}\label{eq_smallparas}
\epsilon \equiv \frac{I_3-I_1}{I_3} \,,\quad
\kappa   \equiv \frac{1}{16}\frac{I_3}{I_1}\frac{I_2-I_1}{I_3-I_2} \,,\quad
\gamma   \equiv \frac{aI_1}{bI_3} \,,
\end{align}
where $\epsilon$ is called the oblateness (or poloidal ellipticity \cite{Wette2023}) of the NS, $\kappa$ describes the ($I_2-I_1$) with respect to the axisymmetric non-sphericity ($I_3-I_2$), while $\gamma$ is called the wobble angle. Their characteristic values are discussed in~\cite{Broeck2005}.
For the small quantities above, 
the expansions of the sines and cosines of the Euler angles (cf. Eqs.~(\ref{eq_Euler_iso})) in $S$-aligned coordinate system up to terms of $O(\gamma^2)$ and $O(\kappa^2)$ are \cite{Gao2020}
\begin{subequations}
\label{eq:Euleranglesexpand}
\begin{align}
\label{eq:costhetaexpand}
\cos{\theta} &= 1-{\gamma ^2}/{2} \,,\quad
\sin{\theta} = \gamma +8 \gamma  \kappa  \sin ^2(t \Omega_{\rm p}) \,,\\
\cos{\phi} &= \cos [t(\Omega_{\rm r}+\Omega_{\rm p})] \,, \,\,
\sin{\phi} = \sin [t(\Omega_{\rm r}+\Omega_{\rm p})] \,,\\
\cos{\psi} &= \sin \left(t \Omega_{\rm p}\right) + 8 \kappa  \sin \left(t \Omega_{\rm p}\right) \cos ^2\left(t \Omega_{\rm p}\right) \\ \nonumber
&+8 \kappa ^2 \left(3 \sin \left(3 t \Omega_{\rm p}\right)-13 \sin \left(t \Omega_{\rm p}\right)\right) \cos ^2\left(t \Omega_{\rm p}\right)  \,,\\ \label{eq:sinpsiexpand}
\sin{\psi} &= \cos \left(t \Omega_{\rm p}\right) -8 \kappa  \sin^2\left(t \Omega_{\rm p}\right) \cos \left(t \Omega_{\rm p}\right) \\ \nonumber
&+ 96 \kappa ^2 \sin ^4\left(t \Omega_{\rm p}\right) \cos \left(t \Omega_{\rm p}\right) \,. 
\end{align}
\end{subequations}

\section{Gravitational waveforms from spinning NS in a binary}
\label{sec:NSinbinarywaveform}

Following the treatment in previous works, such as \cite{LIGONSinBinary2021}, the GWs emitted by the spinning NS in a binary can be obtained by incorporating the Doppler frequency modulation (modulated by orbital precession) into the phase of the triaxial non-aligned waveforms of the NS with spin precession.

First, we calculate the GWs emitted by a spinning NS undergoing spin precession. 
Consider a binary system consisting of a spinning NS with spin angular momentum $\boldsymbol{S}$ and a nonspinning NS (or a slowly spinning NS of which the spin effects can be ignored). This is consistent with the standard evolution scenario of the double NS formed in an isolated system, in which one of the NSs is a rapidly spinning millisecond pulsar and the other is a normal pulsar~\cite{Tauris2017,sl2018}. 
If the companion star has spin, then spin-spin coupling will also cause the rapidly spinning NS to precess. The ratio of the precessional angular frequency due to the spin-spin (SS) coupling to that due to the spin-orbit (SO) coupling satisfies $\Omega_{\rm pre}({\rm {SS}})/\Omega_{\rm pre}({\rm {SO}})<4M/(3M+m_1)(m_2/M)^{1/2}(R_2/r)^{1/2}$, 
where $M=m_1+m_2$ is the total mass of the binary, $m_1$ is the mass of the rapidly spinning NS, $m_2$ and $R_2$ are the mass and the radius of the companion, and $r$ is the orbital separation (cf. Eq. (10.179) in \cite{poisson2014gravity}). For a  double NS system with an orbital period of 10 min, $\Omega_{\rm pre}({\rm {SS}})/\Omega_{\rm pre}({\rm {SO}})<0.01$ for a companion with maximum spin, i.e., the spin-spin precession angular frequency is at least two orders of magnitude smaller than the spin-orbit precession angular frequency, therefore we can ignore the spin-spin coupling. 
Furthermore, in this work we assume that the orbits of binary stars are circular. A recent population synthesis simulation in \cite{Wagg2022} shows that the eccentricities satisfy $e \lesssim 0.01$ for the double NS systems with orbital periods shorter than 10 min. On the other hand, in the extreme cases where eccentricity is important, we need to generalize our current work to incorporate the effects of eccentricity. This can be a subject of our future work.

For a binary system in which only one of the bodies has spin, the precession equations \cite{Apostolatos1994} for the spin of the body and the orbit of the binary
show that to a reasonable approximation the total angular momentum $\boldsymbol J$ maintains its direction, $\boldsymbol S$ keeps its magnitude constant and precesses around $\boldsymbol J$ with
\begin{align}\label{eq:dSdt}
\frac{d \boldsymbol{S}}{d t}= \boldsymbol{\Omega}_{\rm pre} \times \boldsymbol{S} \,,
\end{align}
where 
\begin{equation}\label{eq_Omegapre}
\boldsymbol{\Omega}_{\rm pre} = \frac{G}{2 c^{2} r^{3}}\left(1+\frac{3 M}{m_{1}}\right) \boldsymbol{J} 
\end{equation}
is the angular frequency of spin precession and orbital precession induced by spin-orbit coupling. 
Although the decreasing $r$ and the magnitude of $\boldsymbol J$ due to the radiation reaction cause the magnitude ${\Omega}_{\rm pre}$ of $\boldsymbol{\Omega}_{\rm pre}$ to vary with time, for a typical double NS system with a merger time $\sim O(10^4 {\rm yr})$ and a half-year observation time (see Sec.~\ref{sec:comparisonwithisolated}), the relative variation of ${\Omega}_{\rm pre}$ is $\lesssim 10^{-5}$, so we can assume that ${\Omega}_{\rm pre}$ remains approximately constant in the case considered.

\begin{figure}[!htbp]
\centering
\includegraphics[scale=1.0]{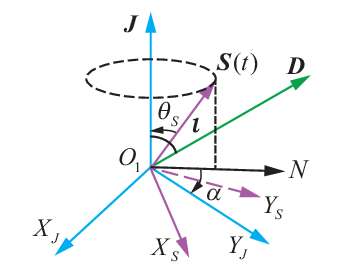}
\caption{$J$-aligned coordinate system $(X_J, Y_J, Z_J)$ with $Z_J$ axis parallel to the total angular momentum $\boldsymbol J$ and $S$-aligned coordinate system $(X_S, Y_S, Z_S)$ with $ Z_S$ axis aligned with spin angular momentum $\boldsymbol{S}$ are both centered at the center of mass ($O_1$) of the spinning NS, where spin-orbit coupling causes $\boldsymbol{S}$ to precess about $\boldsymbol J$ at an angular frequency ${\Omega}_{\rm pre}$. The opening angle of $\boldsymbol{S}$ precession cone is $\theta_S$. Suppose that at initial time, $X_S$ axis coincides with $X_J$ axis, and $\boldsymbol{S}$ is in the $Y_J-Z_J$ plane. The projection of the spin evolution $\boldsymbol{S}(t)$ coincides with $O_{1}N$ in the $X_J-Y_J$ plane, its precession angle $\alpha = \Omega_{\rm pre} t$. The distant observer with position vector $\boldsymbol{D}$ is also assumed in the $Y_J-Z_J$ plane at colatitude $\iota$ from $\boldsymbol J$. }
\label{fig:JS_coord}
\end{figure}

As shown in Fig.~\ref{fig:JS_coord}, $(X_J, Y_J, Z_J)$ is the coordinate system with the origin placed at the center of mass ($O_1$) of the spinning NS and $Z_J$ axis parallel to $\boldsymbol J$ (hereafter
referred to as the $J$-aligned coordinate system), in which the distant observer is assumed to be in the $Y_J-Z_J$ plane with the inclination $\iota$ and the position vector $\boldsymbol D$. The opening angle of $\boldsymbol{S}$ precession cone is $\theta_S$. The coordinate system $(X_S, Y_S, Z_S)$ constructed with $ Z_S$ axis aligned with $\boldsymbol{S}$ is referred to as the $S$-aligned coordinate system. Without loss of generality, we assume that at initial time, the $X_S$ axis coincides with the $X_J$ axis, and $\boldsymbol{S}$ is in the $Y_J-Z_J$ plane. The evolution of $\boldsymbol{S}$ over time is represented by $\boldsymbol{S}(t)$ with the precession angle $\alpha = \Omega_{\rm pre} t$ measured in the $X_J-Y_J$ plane.

To simplify the calculation of waveforms, we use a similar convention for two polarization tensors (cf. Eq.~(\ref{eq_polatensor}) and Eq.~(\ref{eq_transbasis})) as in the calculation of the isolated NS in Sec.~\ref{sec:isolatedNSwaveform}, so that the waveforms in Eqs. (\ref{eq:waveform_h}) also apply to the triaxial non-aligned NS in a binary, but with ($i, R, A$) in Eqs. (\ref{eq:waveform_h}) replaced by ($\iota, \mathcal{R}, \mathcal{A}$), which are the quantities calculated in a coordinate system at rest with respect to the center of mass of the spinning NS, i.e., the $J$-aligned coordinate system shown in Fig~\ref{fig:JS_coord}.
\begin{align}
i \rightarrow \iota , \quad
R \rightarrow \mathcal{R} , \quad
A \rightarrow \mathcal{A} \,.
\end{align}

According to the discussion in \cite{Broeck2005}, the small parameters in Eq.~(\ref{eq_smallparas}) satisfy $\epsilon \ll \kappa \ll \gamma_{\rm max}$ and $\kappa \sim O(\gamma^2)$. Our goal is to expand the CWs of the spinning NS in a tight binary up to order $O(\gamma^2)$ and $O(\kappa)$.
We first calculate $\mathcal{R}$ which represents the rotation matrix from the body coordinate system of the spinning NS to $J$-aligned coordinate system. It can be obtained by the following rotation transformations: first, from the body coordinate system to the $S$-aligned coordinate system by $R$ (cf. Eq.~(\ref{Rmatrix})), and then from the $S$-aligned coordinate system to the $J$-aligned coordinate system by $\mathcal{T}_{S\rightarrow J}$. Thus, 
\begin{equation}\label{eq_newR}
\mathcal{R} = \mathcal{T}_{S\rightarrow J} \cdot R \,,
\end{equation}
where 
\begin{align}\label{RbodytoJ}
&\mathcal{T}_{S\rightarrow J} = \\ \nonumber
&\left(
\begin{array}{ccc}
 \cos (\Omega_{\rm pre} t)  & -\sin (\Omega_{\rm pre} t)  & 0 \\
 \sin (\Omega_{\rm pre} t)  & \cos (\Omega_{\rm pre} t)   & 0 \\
 0 & 0 & 1 \\
\end{array}
\right)
\left(
\begin{array}{ccc}
 1 & 0 & 0 \\
 0 & \cos \theta_S  & \sin \theta_S  \\
 0 & -\sin \theta_S  & \cos \theta_S  \\
\end{array}
\right) \,.
\end{align}
%
After inserting Eqs.~(\ref{eq:Euleranglesexpand}) into Eq.~(\ref{eq_newR}), $\mathcal{R}$ can be expanded as Eqs.~(\ref{eq:Rinbinary}) given in Appendix \ref{App:eq:mathcal_R_and_A}.
Although $\mathcal{R}$ contains $O(\kappa^2)$ terms as in \cite{Gao2020}, we keep them in Eqs.~(\ref{eq:Rinbinary}) in order to facilitate the generalization of the waveforms to higher orders in the future.

Next, we need to calculate $\mathcal{A}$ which depends on the angular frequencies $\omega_i$ of the spinning NS under spin-orbit coupling. The $(1,1)$ and $(1,2)$ components of $\mathcal{A}$ read 
\begin{subequations}
\label{eq_definemathcalA}
\begin{align}
\mathcal{A}_{11} &= 2\left(\Delta_{2} \omega_{2}^{2}-\Delta_{3} \omega_{3}^{2}\right) \,,\\ 
\mathcal{A}_{12} &= \mathcal{A}_{21} = \left(\Delta_{1}-\Delta_{2}\right) \omega_{1} \omega_{2}+\Delta_{3} \dot{\omega}_{3}  \,.
\end{align}
\end{subequations}
The other components of $\mathcal{A}$ are the same as the corresponding ones of $A$ in Eqs.~(\ref{eq:Amatrix}) but with $\Omega_i$ replaced by $\omega_i$.
Since in the body coordinate system $\boldsymbol{S}=S_{\mu}\boldsymbol{e}_{\mu}=I_1 \omega_1 \boldsymbol{e}_1 + I_2 \omega_2 \boldsymbol{e}_2 + I_3 \omega_3 \boldsymbol{e}_3$ 
and 
$\boldsymbol{\Omega}_{\rm pre} = {\Omega}_{\rm pre} \boldsymbol{e}_z = {\Omega}_{\rm pre} \mathcal{R}_{z\mu} \boldsymbol{e}_{\mu}$, then Eq. (\ref{eq:dSdt}) can be expressed as follows 
\begin{equation}
\frac{d{S_{\mu}}}{dt} \boldsymbol{e}_{\mu} + S_{\mu}{\omega}_{\nu}\boldsymbol{e}_{\nu} \times \boldsymbol{e}_{\mu} = {\Omega}_{\rm pre} \mathcal{R}_{z\mu} S_{\nu} \boldsymbol{e}_{\mu} \times \boldsymbol{e}_{\nu} \,,
\end{equation}
with components 
\begin{subequations}
\label{eq:domegaidt}
\begin{align} 
\label{eq:domega1dt}
 \frac{\mathrm{d} \omega_{1}}{\mathrm{d} t} &= \frac{\Delta_{1}}{I_{1}} \omega_{2} \omega_{3} + \frac{\Omega_{\rm pre}}{I_{1}}\left( \mathcal{R}_{z2}I_3\omega_3-\mathcal{R}_{z3}I_2\omega_2 \right)  \,,\\
 \frac{\mathrm{d} \omega_{2}}{\mathrm{d} t} &= \frac{\Delta_{2}}{I_{2}} \omega_{3} \omega_{1} + \frac{\Omega_{\rm pre}}{I_{2}}\left( \mathcal{R}_{z3}I_1\omega_1-\mathcal{R}_{z1}I_3\omega_3 \right)  \,,\\ 
 \label{eq:domega3dt}
 \frac{\mathrm{d} \omega_{3}}{\mathrm{d} t} &= \frac{\Delta_{3}}{I_{3}} \omega_{1} \omega_{2} + \frac{\Omega_{\rm pre}}{I_{3}}\left( \mathcal{R}_{z1}I_2\omega_2-\mathcal{R}_{z2}I_1\omega_1 \right) \,. 
\end{align}
\end{subequations}
The second terms on the right hand sides are the spin-orbit coupling terms.

Although the general solution of the above precession Eqs.~(\ref{eq:domegaidt}) can be obtained numerically, the analytic solution is more favorable for GW detection and parameter estimation because it can be incorporated directly into search algorithms and is more manageable and efficient in data analysis.
We expect the difference between the angular frequency of the spinning NS in an isolated case and that in a binary system to be small (see Fig.~\ref{fig:deltaOmegai}), since the spin-orbit coupling is of 1.5 post-Newtonian order \cite{poisson2014gravity}.
In this sense, we use the perturbation method to solve Eqs. (\ref{eq:domegaidt}) analytically by assuming
\begin{align}\label{eq_pertsol}
\omega_{i} \approx \Omega_{i} + \delta \Omega_{i} \quad (i=1,2,3) \,.
\end{align}
Inserting the above expression into Eqs.~(\ref{eq:domegaidt}) yields the linearized evolution equations 
\begin{widetext}
\begin{equation}\label{linearizeddOmega}
\left(\begin{array}{c}
\frac{d \delta \Omega_{1}}{d t}  \\ [2mm]
\frac{d \delta \Omega_{2}}{d t}  \\ [2mm]
\frac{d \delta \Omega_{3}}{d t} 
\end{array}\right)=\left(\begin{array}{ccc}
0 & \frac{\Delta_1 \Omega_{3}-\Omega_{\rm pre}\mathcal{R}_{z3}I_2}{I_1} & \frac{\Delta_1 \Omega_{2}+\Omega_{\rm pre}\mathcal{R}_{z2}I_3}{I_1} \\ [2mm]
\frac{\Delta_2 \Omega_{3}+\Omega_{\rm pre}\mathcal{R}_{z3}I_1}{I_2} & 0 & \frac{\Delta_2 \Omega_{1}-\Omega_{\rm pre}\mathcal{R}_{z1}I_3}{I_2} \\ [2mm]
\frac{\Delta_3 \Omega_{2}-\Omega_{\rm pre}\mathcal{R}_{z2}I_1}{I_3} & \frac{\Delta_3 \Omega_{1}+\Omega_{\rm pre}\mathcal{R}_{z1}I_2}{I_3} & 0
\end{array}\right)\left(\begin{array}{c}
\delta \Omega_{1} \\ [1mm]
\delta \Omega_{2} \\ [1mm]
\delta \Omega_{3}
\end{array}\right)+\left(\begin{array}{c}
\frac{\Omega_{\rm pre}}{I_{1}}\left( \mathcal{R}_{z2}I_3\Omega_3-\mathcal{R}_{z3}I_2\Omega_2 \right) \\ [2mm]
\frac{\Omega_{\rm pre}}{I_{2}}\left( \mathcal{R}_{z3}I_1\Omega_1-\mathcal{R}_{z1}I_3\Omega_3 \right) \\ [2mm]
\frac{\Omega_{\rm pre}}{I_{3}}\left( \mathcal{R}_{z1}I_2\Omega_2-\mathcal{R}_{z2}I_1\Omega_1 \right)
\end{array}\right)  \,.
\end{equation}
\end{widetext}

According to Appendix \ref{App:residual}, $\delta \Omega_3 \ll \delta \Omega_{1,2}$, so the contribution of $\delta \Omega_3$ can be ignored when solving for $\delta \Omega_{1,2}$. 
For a typical double NS system with 6 min orbital period discussed below, $\Omega_{\rm pre}/{\rm Hz} \lesssim \epsilon \ll \kappa \lesssim \gamma^2$. 
We first expand $\Omega_i$ (cf. Eqs.~(\ref{eq_iso_Omega})) to leading order. The combination of Eq.~(\ref{eq_tau}), Eq.~(\ref{eq_T}), and Eq.~(\ref{eq_preandrotfreq}) gives $\tau=(2K(m)/{\pi})t\Omega_{\rm p}=(1+O(m))t\Omega_{\rm p}$. From Eq.~(\ref{eq_m}) and Eq.~(\ref{eq_smallparas}), we have $m=16\gamma^2\kappa$. Therefore, $\tau \simeq t\Omega_{\rm p}$ to leading order. In a similar way, to leading order, the Jacobian elliptic functions become ${\rm cn}(\tau, m) \simeq \cos(t\Omega_{\rm p})$, ${\rm sn}(\tau, m) \simeq \sin(t\Omega_{\rm p})$, and ${\rm dn}(\tau, m) \simeq 1$, and the factor $\left[{I_{1}\left(I_{3}-I_{1}\right)}/{I_{2}/\left(I_{3}-I_{2}\right)}\right]^{1 / 2} \simeq 1$. Consequently, we have the leading order result
\begin{subequations}\label{eq_expand_iso_Omega}
\begin{align}
\Omega_{1} &= a \cos(t\Omega_{\rm p}) \,, \\
\Omega_{2} &= a \sin(t\Omega_{\rm p}) \,, \\ 
\Omega_{3} &= b   \,.
\end{align}
\end{subequations}
By inserting Eqs.~(\ref{eq_expand_iso_Omega}) and Eqs.~(\ref{eq:Rinbinary}) into Eq.~(\ref{linearizeddOmega}), we expand the linearized evolution equations up to order $O(\gamma^2)$ and $O(\kappa)$, and ignore higher order terms such as $O(\gamma^2 \kappa)$, $O(\gamma \Omega_{\rm pre})$ and $O(\kappa \Omega_{\rm pre})$. To give analytic approximations for $\delta \Omega_i$, we further simplify the equations by setting $1+\epsilon \simeq 1$ and discard $O(\Omega_{\rm pre})$ in $O(\Omega_{\rm pre}) + O(b \epsilon)$ due to $O(b) \sim {2\pi} \times 100~\rm{Hz}$. Finally, we obtain
\begin{subequations}
\begin{align}
\frac{d \delta \Omega_{1}}{d t} &\simeq b \epsilon (-1+16\kappa) \delta \Omega_2 + b \Omega_{\rm pre} \sin \theta_{\rm S} \sin(t\Omega_{\rm r})   \,, \\
\frac{d \delta \Omega_{2}}{d t} &\simeq b \epsilon \delta \Omega_1 + b \Omega_{\rm pre} \sin \theta_{\rm S} \cos(t\Omega_{\rm r})  \,.
\end{align}\label{eq:pert}
\end{subequations}
Setting the initial conditions $(\delta \Omega_1(0),\delta \Omega_2(0),\delta \Omega_3(0)) = (0,0,0)$, and ignoring $O(b \epsilon)$ in  $O(b \epsilon) + O(\Omega_{\rm r})$ due to $b \simeq \Omega_{\rm r}$,
the integration of Eqs.~(\ref{eq:pert}) yields 
\begin{subequations}\label{eq_dOmega12}
\begin{align}
\delta \Omega_1 &\simeq {\Omega_{\rm pre} \sin \theta_{\rm S} \left[\cos (t b\epsilon \sqrt{1-16\kappa})-\cos (t \Omega_{\rm r})\right]}  \,, \\
\delta \Omega_2 &\simeq {\Omega_{\rm pre} \sin \theta_{\rm S} \left[\sin (t b\epsilon \sqrt{1-16\kappa})+\sin (t \Omega_{\rm r})\right]}  \,.
\end{align}
\end{subequations}
Using Eqs.~(\ref{eq_dOmega12}), the integration of the third row in Eq.~(\ref{linearizeddOmega}) leads to
\begin{equation}\label{eq_dOmega3}
\delta \Omega_3 \simeq { \gamma \Omega_{\rm pre}\sin{\theta_S} [\cos (t(\Omega _{\rm p}+\Omega_{\rm r}))-1]} \,. 
\end{equation}
The error of this analytic approximation is shown and discussed in Appendix \ref{App:residual}.

Now, $\mathcal{A}$ can be calculated by inserting $\Omega_i$ (cf. Eqs.~(\ref{eq_expand_iso_Omega})) and $\delta \Omega_i$ (cf. Eqs.~(\ref{eq_dOmega12}) and Eq.~(\ref{eq_dOmega3})) into Eqs.~(\ref{eq_definemathcalA}). The result is given in Eqs. (\ref{eq:Amatrixinbinary}) of Appendix \ref{App:eq:mathcal_R_and_A}.

With $\mathcal{R}$ and $\mathcal{A}$, the waveforms emitted by the spinning NS undergoing spin precession can be expressed in terms of the series expansion as follows 
\begin{subequations}
\begin{align}
h_+ &= h_+^{(1)} + h_+^{(2)} + h_+^{(3)} + \cdots \\
h_\times &= h_\times^{(1)} + h_\times^{(2)} +  h_\times^{(3)} + \cdots
\end{align}
\end{subequations}
where
\begin{widetext}
\begin{subequations}
\begin{align}\label{eq:h1p}
h_+^{(1)} &= \frac{{G{b^2}{I_3}\epsilon \gamma }}{{4{c^4}D}}\Big[\cos(t({\Omega _{\rm{p}}} + {\Omega _{\rm{r}}}))[\sin (2{\theta _S})( 6{\sin ^2}\iota - (3 + \cos (2\iota ))\cos (2t{\Omega _{{\rm{pre}}}}))+ 4\cos (2{\theta _S})\cos (t{\Omega _{{\rm{pre}}}})\sin (2\iota )] \\ \nonumber
& + 2[ - 2\cos {\theta _S}\sin (2\iota )\sin (t{\Omega _{{\rm{pre}}}})+ (3 + \cos (2\iota ))\sin {\theta _S}\sin (2t{\Omega _{{\rm{pre}}}})]\sin (t({\Omega _{\rm{p}}} + {\Omega _{\rm{r}}}))\Big] \,, \\
h_ \times ^{(1)} &= \frac{{G{b^2}{I_3}\epsilon \gamma }}{{{c^4}D}}\Big[\cos (t({\Omega _{\rm{p}}} + {\Omega _{\rm{r}}}))[2\sin \iota \cos (2{\theta _S})\sin (t{\Omega _{{\rm{pre}}}}) - \cos \iota \sin(2{\theta _S})\sin (2t{\Omega _{{\rm{pre}}}})] \\ \nonumber
 & + 2[\sin \iota \cos {\theta _S}\cos (t{\Omega _{{\rm{pre}}}})- \cos \iota \cos (2t{\Omega _{{\rm{pre}}}})\sin {\theta _S}]\sin (t({\Omega _{\rm{p}}} + {\Omega _{\rm{r}}}))\Big] \,, \\
h_ + ^{(2)} &=  - \frac{{16G{b^2}{I_3}\epsilon \kappa }}{{{c^4}D}}\Big[(3 + \cos (2\iota ))[\cos^4 \left( {\frac{{{\theta _S}}}{2}} \right)\cos (2t({\Omega _{{\rm{pre}}}} + {\Omega _{\rm{r}}}))+ \cos (2t({\Omega _{{\rm{pre}}}} - {\Omega _{\rm{r}}})){\sin ^4}\left( {\frac{{{\theta _S}}}{2}} \right)] \\ \nonumber
 &+ \cos (2t{\Omega _{\rm{r}}})[3{\sin ^2}{\theta _S}{\sin ^2}\iota  + \cos (t{\Omega _{{\rm{pre}}}})\sin(2{\theta _S})\sin (2\iota )]- 2\sin {\theta _S}\sin (2\iota )\sin (t{\Omega _{{\rm{pre}}}})\sin (2t{\Omega _{\rm{r}}})\Big] \,, \\ \label{eq:h2c}
h_ \times ^{(2)} &= \frac{{16G{b^2}{I_3}\epsilon \kappa }}{{{c^4}D}}\Big[\cos (2t{\Omega _{\rm{r}}})[ - 2\sin \iota \sin (2{\theta _S})\sin (t{\Omega _{{\rm{pre}}}})- \cos \iota (3 + \cos (2{\theta _S}))\sin (2t{\Omega _{{\rm{pre}}}})]\\ \nonumber
 &- 4[\cos \iota \cos {\theta _S}\cos (2t{\Omega _{{\rm{pre}}}})+ \sin \iota \cos (t{\Omega _{{\rm{pre}}}})\sin {\theta _S}]\sin (2t{\Omega _{\rm{r}}})\Big] \,, \\
 \label{eq:h3p}
h_ + ^{(3)} &= \frac{Gb^2{I_3}\epsilon \gamma^2}{c^4D}\Big[(3 + \cos (2\iota ))[\cos^4 \left( {\frac{{{\theta _S}}}{2}} \right)\cos (2t({\Omega _{{\rm{pre}}}} + {\Omega _{\rm{r}}}+\Omega _{\rm{p}}))+ \cos (2t({\Omega _{{\rm{pre}}}} - {\Omega _{\rm{r}}}-\Omega _{\rm{p}})){\sin ^4}\left( {\frac{{{\theta _S}}}{2}} \right)] \\ \nonumber
 &+ \cos (2t(\Omega _{\rm{r}}+\Omega _{\rm{p}}))[3{\sin ^2}{\theta _S}{\sin ^2}\iota  + \cos (t{\Omega _{{\rm{pre}}}})\sin(2{\theta _S})\sin (2\iota )]- 2\sin {\theta _S}\sin (2\iota )\sin (t{\Omega _{{\rm{pre}}}})\sin (2t(\Omega _{\rm{r}}+\Omega _{\rm{p}}))\Big] \,, \\ \label{eq:h3c}
h_{\times} ^{(3)} &= \frac{Gb^2{I_3}\epsilon \gamma^2}{{{c^4}D}}\Big[\cos (2t(\Omega _{\rm{p}}+\Omega _{\rm{r}}))[2\sin (2{\theta _S})\sin \iota \sin (t{\Omega _{{\rm{pre}}}})+ (3 + \cos (2\theta _S)) \cos \iota \sin (2t{\Omega _{{\rm{pre}}}})]  \\ \nonumber 
 & + 4[\cos {\theta _S}\cos (2t{\Omega _{{\rm{pre}}}})\cos \iota +\cos (t\Omega _{\rm{pre}}) \sin{\theta_S} \sin{\iota}] \sin (2t(\Omega _{\rm{p}}+\Omega _{\rm{r}}))\Big] \,.
\end{align}\label{eq:hinbinary}
\end{subequations}
\end{widetext}
Note that $\iota$ in the above expressions is the angle between $\boldsymbol{J}$ and the line of sight, not to be confused with the inclination defined by the angle between $\boldsymbol{S}$ and the line of sight for the isolated case. 

Here, $h_{+,\times}^{(1)}$ are of order $O(\gamma)$, $h_{+,\times}^{(2)}$ are of order $O(\kappa)$, $h_{+,\times}^{(3)}$ are of order $O(\gamma^2)$.
Note that only the waveform components up to terms of $O(\gamma)$,  $O(\kappa)$ and $O(\gamma^2)$ are shown here, according to the discussion of the characteristic value of $\gamma$ and $\kappa$ in \cite{Broeck2005}. The $O\left(\Omega_{\rm pre}\right)$ components (see Appendix \ref{App:eq:h+xpre}), and higher-order $O(\gamma \kappa)$, and $O(\kappa^2)$ components can be ignored for the parameter values we adopt in the following analysis. 

Then, the waveforms of the spinning NS in a binary system with spin-orbit coupling effects considered are completed by incorporating the Doppler frequency modulation of this NS around the binary barycenter (BB) into the phases of waveforms Eqs.~(\ref{eq:hinbinary}), which is done by the second term on the right-hand side of Eq.~(\ref{eq:Dopshift}) in Sec.~\ref{sec:detection}.
We leave this for further discussion in Sec.~\ref{sec:detection}, where this Doppler modulation and the Doppler modulation due to the motion of the GW detector around the solar system barycenter (SSB) are combined, as in \cite{LIGONSinBinary2021}.

\section{Comparison with waveforms from isolated NS}
\label{sec:comparisonwithisolated}

As a limiting case, when the spinning NS is isolated and spin $\boldsymbol{S}$ is along the $z$-axis of the coordinate system, i.e., $\Omega_{\rm pre}=\theta_S=0$, the waveforms in Eqs.~(\ref{eq:hinbinary}) will reduce to those given in \cite{Gao2020}:
\begin{subequations}\label{eq_isoNS}
\begin{align}
h_{+}^{(1)} &= \frac{G b^2 I_3 \epsilon \gamma  \sin (2 \iota ) \cos (t (\Omega_{\rm p}+\Omega_{\rm r}))}{c^4 D} \,, \\
h_{\times}^{(1)} &= \frac{2 G b^2 I_3 \epsilon \gamma  \sin \iota  \sin (t (\Omega_{\rm p}+\Omega_{\rm r}))}{c^4 D} \,, \\
h_{+}^{(2)} &= -\frac{32 G b^2 I_3 \epsilon  \kappa  \left(1+\cos^2\iota\right)  \cos (2 t\Omega_{\rm r})}{c^4 D} \,, \\
h_{\times}^{(2)} &= -\frac{64 G b^2 I_3 \epsilon  \kappa  \cos\iota  \sin (2 t\Omega_{\rm r})}{c^4 D} \,, \\
h_{+}^{(3)} &= \frac{2 G b^2 I_3 \epsilon \gamma^2  \left(1+\cos^2\iota\right) \cos (2t (\Omega_{\rm p}+\Omega_{\rm r}))}{c^4 D} \,, \\
h_{\times}^{(3)} &= \frac{4 G b^2 I_3 \epsilon \gamma^2  \cos \iota  \sin (2t (\Omega_{\rm p}+\Omega_{\rm r}))}{c^4 D} \,.
\end{align}
\end{subequations}
Note that some of the signs in the above equations are reversed because the convention for the inclination ($\iota$) we use here \cite{Zimmermann1980} is equivalent to  ($\pi-\iota$) in \cite{Gao2020}.

For small equatorial ellipticity $\varepsilon \equiv |I_1-I_2|/I_3 \simeq 16 \epsilon \kappa$ and $\gamma = 0$, $h_{+,\times}^{(1)}=h_{+,\times}^{(3)}=0$ and $h_{+,\times}^{(2)}$ will reduce to the triaxial aligned waveforms (e.g., Eq. (4.223) of \cite{Maggiore2007}). 
If $\kappa=0$ and $\gamma \ne 0$, $h_{+,\times}^{(2)}=0$,  $h_{+,\times}^{(1)}$ and $h_{+,\times}^{(3)}$ will reduce to the biaxial waveforms (cf. Eq. (1) of \cite{Zimmermann1979}).

The values of the parameters used in the following analysis are intended to make the signal amplitude as large as possible under current observational and theoretical constraints. 
The NSs, measured by pulsar timing \cite{Demorest2010Natur} or GW observation \cite{GW190425}, can have masses up to about two solar masses. 
The widely accepted range of the moment of inertia for NSs resides in $1-3 \times 10^{38}~{\rm kg~m^2}$ (see \cite{Lu2022arXiv} and references therein). 
The parameters that characterize the properties of the spinning NS are in accordance with the discussion in \cite{Broeck2005}, 
in which $\epsilon \ll \kappa \ll \gamma_{\rm max}$ and $\kappa \sim O(\gamma^2)$.
Recent observations from Advanced LIGO and Advanced Virgo constrain two recycled pulsars (PSR J0437-4715 and PSR J0711-6830) to have equatorial ellipticities ($\varepsilon \equiv |I_1-I_2|/I_3 \simeq 16 \epsilon \kappa$) of less than $10^{-8}$ \cite{Ellipticity2020ApJ}. An ellipticity of $10^{-8}$ is also a typical value used in searching for CWs from small-ellipticity sources \cite{Dergachev2020PhRvL}.
Population syntheses of Galactic disk double NS systems detectable by LISA~\cite{Lau2020, Andrews2020, Wagg2022} and TianQin~\cite{Feng2023} suggest that the existence of double NSs with orbital periods as low as 6 minutes. We assume that the orbital period under consideration is 6 minutes, which maximizes the strength of the orbital precession.
Based on the current observations of 22 double NSs \cite{ATNF_website, Manchester2005}, we select $10~{\rm ms}$ as the typical spin period of the rapidly spinning NS \cite{Feng2023}, and $1~{\rm kpc}$ as the typical distance since about half of the known double NSs are located near that value \cite{ATNF_website, Manchester2005}.

\begin{figure}[!htbp]
\centering
\includegraphics[scale=0.62]{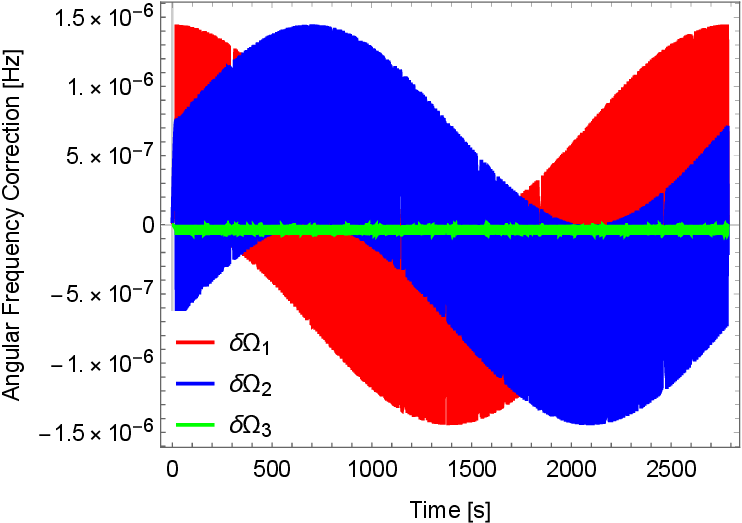}
\caption{The analytic approximate solution for $\delta\Omega_i$ in Eqs.~(\ref{eq_dOmega12}) and (\ref{eq_dOmega3}) during a free precession period ($T=2\pi/\Omega_{\rm p}=2785~{\rm s}$). Assume that the binary mass $m_1 = m_2 = 2.0~M_{\odot}$, the orbital period $P_{\rm b}=6~{\rm min}$, and the opening angle of the spin precession cone is $\theta_{\rm S}=5\pi/12$. The spinning NS's characteristic parameters are $I_3=2.0\times 10^{38}~{\rm kg~m^2}$, $\epsilon=3.6\times 10^{-6}$, $\kappa=1.75\times 10^{-4}$, $\gamma=5.0\times 10^{-2}$ (equatorial ellipticity $\varepsilon \equiv |I_1-I_2|/I_3 =1.0\times 10^{-8}$). $\delta \Omega_{1,2}$ ($\delta \Omega_{3}$) can reach up to $1.45\times 10^{-6}~{\rm Hz}$ ($3.62\times 10^{-8}~{\rm Hz}$) for $\Omega_{\rm r}=628.32~{\rm Hz}$, $\Omega_{\rm p}=2.26\times 10^{-3}~{\rm Hz}$, $\Omega_{\rm pre}=7.50\times 10^{-7}~{\rm Hz}$ obtained with the above values of the parameters. }
\label{fig:deltaOmegai}
\end{figure}

As an example, we assume the component masses of a double NS system $m_1 = m_2 = 2.0~M_{\odot}$, the orbital period $P_{\rm b}=6~{\rm min}$, and the opening angle of $\boldsymbol{S}$ precession cone is $\theta_{\rm S}=5\pi/12$. 
The spinning NS's characteristic parameters are $I_3=2.0\times 10^{38}~{\rm kg~m^2}$, $\epsilon=3.6\times 10^{-6}$, $\kappa=1.75\times 10^{-4}$, $\gamma=5.0\times 10^{-2}$ (equatorial ellipticity $\varepsilon \equiv |I_1-I_2|/I_3 = 1.0\times 10^{-8}$). 
From Eq.~(\ref{eq_Omegapre}) and Eq.~(\ref{eq_preandrotfreq}), one can obtain $\Omega_{\rm r}=628.32~{\rm Hz}$, $\Omega_{\rm p}=2.26\times 10^{-3}~{\rm Hz}$, and $\Omega_{\rm pre}=7.50\times 10^{-7}~{\rm Hz}$.
During a free precession period ($T=2\pi/\Omega_{\rm p}=2785~{\rm s}$), Fig. \ref{fig:deltaOmegai} shows the approximate solution for $\delta\Omega_i$ in Eqs.~(\ref{eq_dOmega12}) and (\ref{eq_dOmega3}), which can reach up to $1.45\times 10^{-6}~{\rm Hz}$ for $\delta \Omega_{1,2}$ and $3.62\times 10^{-8}~{\rm Hz}$ for $\delta \Omega_{3}$.
Note that the jagged profiles in this figure are due to a reduced sampling rate over a long spin precession period, as in the following figures.

\begin{figure*}[!htbp]
\centering
\includegraphics[scale=0.27]{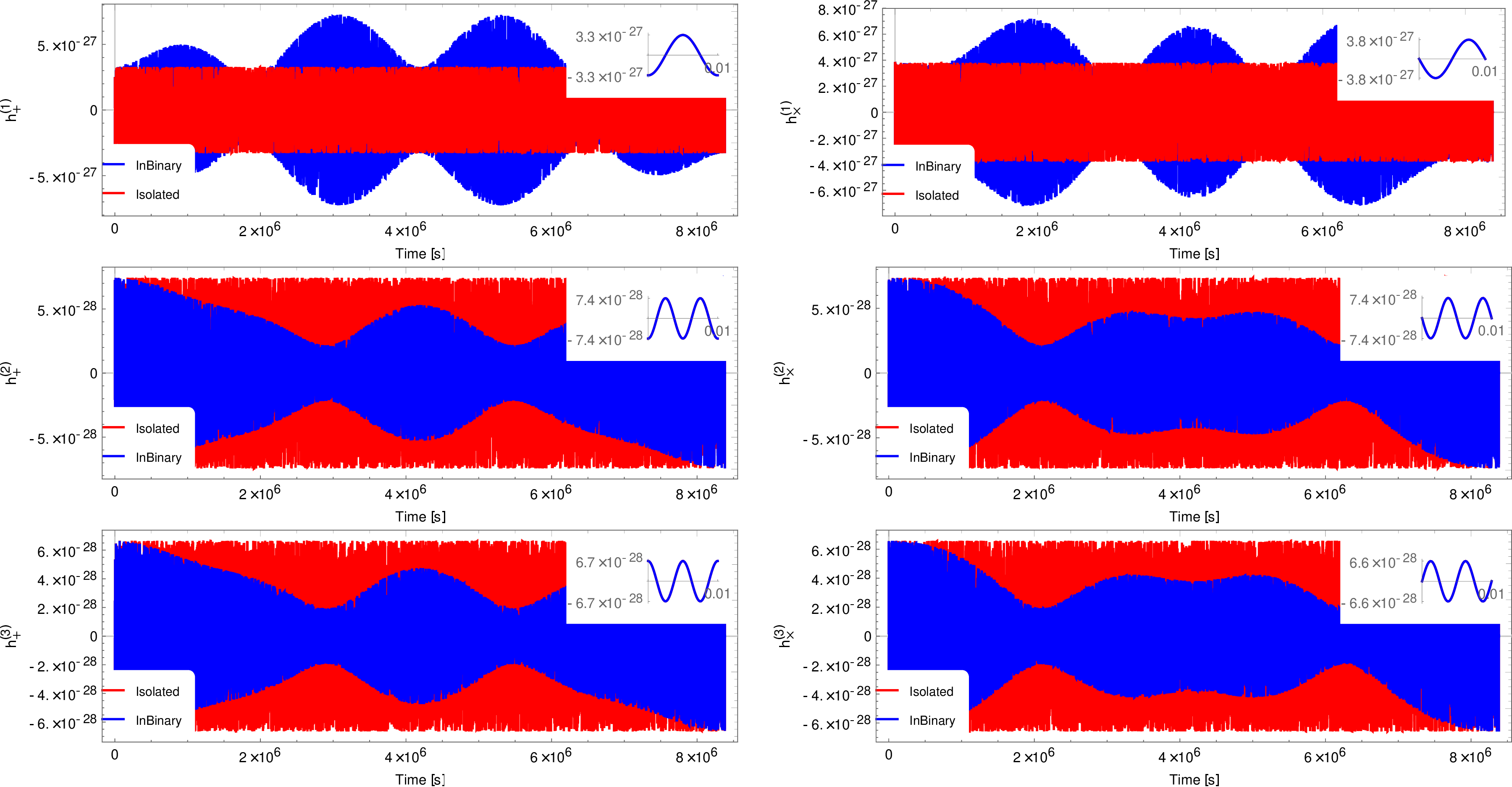}
\caption{The waveform components from the triaxial non-aligned NS with spin precession incorporated (blue) during a spin precession period and in comparison with those from the isolated NS (red). The values of the parameters used here are the same as in Fig.~\ref{fig:deltaOmegai}, and set the spin period of the NS $P_{\rm s}=10~{\rm ms}$, the inclination angle of $\boldsymbol{J}$ with respect to the line of sight $\iota=\pi/4$, and the distance to the observer $D=1~{\rm kpc}$. The modulated amplitude profiles in a binary depend on $\iota$, $\theta_S$, and $\Omega_{\rm pre}$, which can be several times larger or smaller than the isolated case for different waveform components. The inset in each subplot shows the waveform during the first spin period of the NS.
}
\label{fig:h_inbinary_isolated}
\end{figure*}

Fig. \ref{fig:h_inbinary_isolated} shows the different waveform components with spin precession incorporated during a spin precession period ($T_{\rm pre}=2\pi/\Omega_{\rm pre}=8.382 \times 10^6~{\rm s}$). The values of the parameters used here are the same as in Fig. \ref{fig:deltaOmegai}, and we set the spin period of the NS $P_{\rm s}=10~{\rm ms}$, the inclination angle of $\boldsymbol{J}$ with respect to the line of sight $\iota=\pi/4$, and the distance to the observer $D=1~{\rm kpc}$. 
The modulated amplitude profiles for a binary depend on $\iota$, $\theta_S$, and $\Omega_{\rm pre}$, which can be several times larger or smaller than the isolated case for different waveform components.
The amplitude modulations shown here are only due to the NS's spin precession caused by spin-orbit coupling, the Doppler modulation will be included in Sec.~\ref{sec:detection} (see the discussion at the end of Sec.~\ref{sec:NSinbinarywaveform}).

According to Eqs.~(\ref{eq_isoNS}), the GW angular frequency components of an isolated NS are $\Omega_{+,\times}^{\rm iso} = \Omega_{\rm r}+\Omega_{\rm p}$, $2\Omega_{\rm r}$, and $2(\Omega_{\rm r}+\Omega_{\rm p})$ for both $+$ and $\times$ polarizations. 
Spin-orbit coupling in the binary can split these frequencies in the following way 
\begin{subequations}\label{eq:splitfreq}
\begin{align}
\Omega_{+}^{\rm iso} &\longrightarrow \Omega_{+}^{\rm iso} \pm n \Omega_{\rm pre} ~(n=0,1,2)  \,, \\
\Omega_{\times}^{\rm iso} &\longrightarrow \Omega_{\times}^{\rm iso} \pm n \Omega_{\rm pre} ~(n=1,2)  \,.  
\end{align}
\end{subequations}
Similar to the analysis of the GW spectrum of isolated systems \cite{Broeck2005}, the spectral analysis of the above frequency components can be used to infer the orbital period of the binary and the characteristic parameters of the NS. 

\begin{figure}[!htbp]
\centering
\includegraphics[scale=0.64]{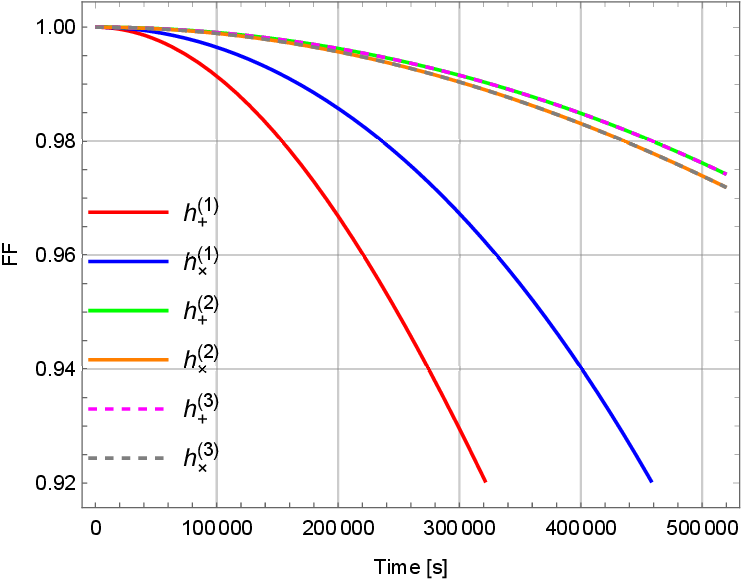}
\caption{The fitting factors between the isolated case and the one with spin precession for different waveform components shown in Fig. \ref{fig:h_inbinary_isolated}. }
\label{fig:FF_Time}
\end{figure}

In order to quantitatively measure the degree of matching between these two types of waveforms, one can calculate the fitting factor (FF) \cite{Apostolatos1995FF} between the genuine GW waveforms generated by the spinning NS in a binary (denoted as $h_{\rm b}$) and the ones by an isolated NS (denoted as $h_{\rm i}$), given the latter has been used in the matched filtering of CW data analysis 
\begin{equation}
{\rm{FF}} \equiv \max _{\boldsymbol {\lambda}} \frac{\left(h_{\rm b}, h_{\rm i}\right)}{\sqrt{\left(h_{\rm b}, h_{\rm b}\right)(h_{\rm i}, h_{\rm i})}} \,,
\end{equation}
where ${\boldsymbol{\lambda}}$ is the set of the parameters that characterize the waveforms. For a quasi-monochromatic signal, the inner product $\left(h_{\rm b}, h_{\rm i}\right)$ can be simplified as $ \left(h_{\rm b}, h_{\rm i}\right) \equiv \int_{0}^{T_{\rm obs}} h_{\rm b}(t) h_{\rm i}(t) dt  \,, $
where $T_{\rm obs}$ is the observation time. 
Fig.~\ref{fig:FF_Time} shows the FFs of different waveform components in $h_{\rm b}$ and $h_{\rm i}$ as a function of $T_{\rm obs}$. 
As we can see, the two waveforms only match well (${\rm FF}>0.97$) within roughly a few days ($\sim O(10^5~{\rm s})$) and then start to diverge rapidly. 

\section{Detecting GW from spinning NS in a binary}
\label{sec:detection}

\subsection{Signal model in detector coordinate system}
\label{subsec_signalmodel}

The GW strain signal from a spinning triaxial non-aligned NS in a binary can be written as a sum of different waveform components $h_n(t)~(n=1,2,3)$ as follows \cite{Jaranowski1998}
\begin{align}\label{eq:ht}
h(t) = \sum_{n=1}^3 h_n(t) = \sum_{n=1}^3 (F_{+}(t) H_{+}^{(n)}(t) + F_{\times}(t) H_{\times}^{(n)}(t) )  \,,
\end{align}
where 
$H_{+,\times}^{(n)}(t)$ are the Doppler-modulated waveforms (cf. Eqs.~(\ref{eq_includeDop}) below),
and $F_{+,\times}(t)$ are the antenna pattern functions of GW detector, of which the explicit expressions can be found in \cite{Jaranowski1998}. $F_{+,\times}(t)$ depend on a set of parameters listed below:
$\gamma_{\rm o}$ characterizes the orientation of the detector with respect to the local geographical directions, $\zeta$ denotes the angle between the interferometer arms, $\lambda$ is the geographical latitude of the detector's site, $(\alpha,\delta)$ are the right ascension and declination of the source, $\psi_{\rm p}$ is the GW polarization angle, $\Omega_{\rm Er}$ is the rotational angular frequency of the Earth, and $\phi_{\rm r}$ is the initial phase of the Earth's diurnal motion.

\begin{figure}[!htbp]
\centering
\includegraphics[scale=0.9]{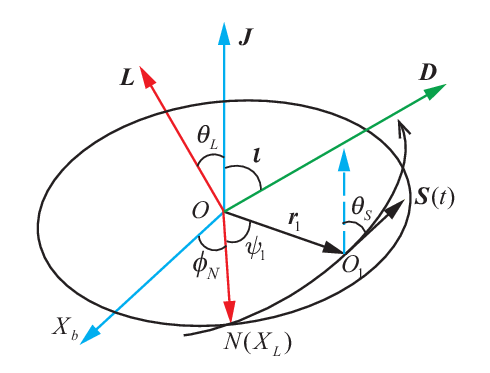}
\caption{The binary coordinate system $(X_b, Y_b, Z_b)$ and the motion of the spinning NS within it. $(X_L, Y_L, Z_L)$ is the coordinate system with $Z_L$ axis aligned with the binary's orbital angular momentum $\boldsymbol L$, $X_L-Y_L$ represents the orbital plane. The opening angle of $\boldsymbol L$ precession cone is $\theta_L$. The distant detector with position vector $\boldsymbol{D}$ is assumed in the $Y_b-Z_b$ plane at colatitude $\iota$ from $\boldsymbol J$. }
\label{fig:JLS_coord}
\end{figure}

Below we will consider how to incorporate Doppler modulation into the phases of the waveforms from the spinning NS undergoing spin precession (cf. Eqs.~(\ref{eq:hinbinary})). 
Fig.~\ref{fig:JLS_coord} shows the binary coordinate system $(X_b, Y_b, Z_b)$ with $Z_b$ axis aligned with the binary's total angular momentum $\boldsymbol J$ and the motion of spinning NS within it. $(X_L, Y_L, Z_L)$ is the coordinate system with $Z_L$ axis aligned with the binary's orbital angular momentum $\boldsymbol L$ (hereafter referred to as $L$-aligned coordinate system), in which $X_L-Y_L$ represents the orbital plane. The origins of these two frames are both placed at the BB ($O$). The opening angle of $\boldsymbol L$ precession cone is $\theta_L$.
Suppose that at initial time, $\boldsymbol{S}$ and $\boldsymbol{L}$ are both in the $Y_b-Z_b$ plane and the spinning NS sits on the $X_b$ axis. After a period of time $t$, the orbital plane precessed by $\phi_{\rm N}$, the spinning NS's position vector and the orbital longitude are $\boldsymbol{r}_1$ and $\psi_1$, and its spin is $\boldsymbol{S}(t)$. (The $Y$-axes not drawn in Fig.~\ref{fig:JLS_coord} are determined by the right-hand rule.)

In the $L$-aligned coordinate system, the position vector of the spinning NS $\boldsymbol{r}_{1L} = (r_1 \cos(\omega_{\rm b}t),r_1 \sin(\omega_{\rm b}t),0)$. 
By two rotation transformations (first rotates by $\theta_L$ clockwise about the $X_L$ axis, and then rotates by $\phi_N$ clockwise about the $Z_b$ axis), the position vector $\boldsymbol{r}_{1}$ in the binary coordinate system can be given by 
\begin{align}\label{rL2rJ}
\boldsymbol{r}_{1} = r_1 \left(
\begin{array}{c}
\cos \psi _1 \cos \phi _N - \cos \theta _L \sin \psi _1 \sin \phi _N  \\
\cos \theta _L \sin \psi _1 \cos \phi _N + \cos \psi _1 \sin \phi _N  \\
\sin \theta _L \sin \psi _1  \, \\
\end{array}
\right)
\end{align}
with $\psi _1=\omega_{\rm b}t$ and $\phi _N=\Omega_{\rm pre}t$.

In the detector coordinate system, the Doppler shift of the GW frequency $f_0$ from the spinning NS in a binary system can be expressed as the combination of detector Doppler shift around the SSB and source Doppler shift around the BB 
\begin{equation}\label{eq:Dopshift}
\Delta f_{\rm{D}} \approx f_0 \left(\frac{{\boldsymbol n} \cdot {\rm d}\boldsymbol{r}_{\rm d}/{\rm d}t}{c}+\frac{{\boldsymbol n}_{\rm b} \cdot {\rm d}\boldsymbol{r}_1/{\rm d}t}{c}\right)  \,,
\end{equation}
with \cite{Jaranowski1998}
\begin{align}
\boldsymbol{n} \cdot \boldsymbol{r}_{\rm d} &= R_{\rm{E}}[\cos \lambda \cos \delta \cos (\alpha-\phi_r-\Omega_{\rm Er} t) \\ \nonumber
&+\sin \lambda \sin \delta] +R_{\rm{ES}}[\cos \alpha \cos \delta \cos (\phi_{\rm o}+\Omega_{\rm Eo} t) \\ \nonumber
&+(\cos \varepsilon_{\rm e} \sin \alpha \cos \delta+\sin \varepsilon_{\rm e} \sin \delta) \sin (\phi_{\rm o}+\Omega_{\rm Eo} t) ]  \nonumber
\end{align}
being the projection of the detector's position vector $\boldsymbol{r}_{\rm d}$ along the spinning NS's line of sight $\boldsymbol{n}$ in SSB coordinate system, where
$R_{\rm E}$ and $R_{\rm ES}$ are the mean radius of the Earth and the mean distance from the Earth’s center to the SSB, $\Omega_{\rm Eo}$ is the mean orbital angular frequency of the Earth, $\phi_{\rm o}$ is the initial phase of the Earth's annual motion, and $\varepsilon_{\rm e}$ is the ecliptic obliquity.
$-{\boldsymbol n}_{\rm b}=(0,\sin{\iota},\cos{\iota})$ is the SSB's location in the binary coordinate system, orbital radius $r_1=G^{1/3}m_2/(\omega_{\rm b}M)^{2/3}$ and angular frequency $\omega_{\rm b} = 2\pi /P_{\rm b}$ for a circular orbit.
If there is no orbital precession, i.e., $\Omega_{\rm pre}=0$, then the Doppler shift in Eq.~(\ref{eq:Dopshift}) reduces to the simple case as Eq. (6) in \cite{LIGONSinBinary2021}.

Since the waveform components in Eqs.~(\ref{eq:hinbinary}) can be decomposed into a series of sine and cosine functions in which the frequencies are linear combinations of $\Omega_{\rm r}$, $\Omega_{\rm p}$, and $\Omega_{\rm pre}$, the Doppler-modulated waveforms $H_{+,\times}^{(n)}(t)$ in Eq.~(\ref{eq:ht}) can be obtained by
\begin{subequations}\label{eq_includeDop}
\begin{align} 
H_{+}^{(n)}(t) &= h_{+}^{(n)}(\Omega_{l} \rightarrow \Omega_{l} + 2\pi \Delta f_{\rm{D}}) \,, \\
H_{\times}^{(n)}(t) &= h_{\times}^{(n)}(\Omega_{l} \rightarrow \Omega_{l} + 2\pi \Delta f_{\rm{D}}) \,, 
\end{align}
\end{subequations}
with $l\in \{\rm r, p, pre\}$.

The binary is assumed to follow an invariant circular orbit when calculating the Doppler shift of the spinning NS. In fact, the orbit is constantly shrinking due to gravitational radiation. According to the orbital velocity evolution equation ($d(v/c)/dt \propto (v/c)^9$, cf. \cite{Creighton2011}), its relative variation is $\lesssim 10^{-5}$ for a half-year observation, resulting in a relative variation for the Doppler shift of $\lesssim 10^{-5}$. Thus, we can ignore the reaction of gravitational radiation on the orbit for the observation time under consideration.

\subsection{Effects of spin-orbit coupling on parameter estimation}

In addition to spin precession, spin-orbit coupling also causes orbital plane precession, which is expected to introduce additional information into the Doppler-modulated waveforms $H_{+,\times}^{(n)}(t)$ (cf. Eqs.~(\ref{eq_includeDop})).
We use the Fisher information matrix (FIM) to obtain a quantitative assessment of the parameter estimation accuracy for GW detection (e.g., see \cite{Jaranowski1999PhRvD}). 
For GW signal $h(t)$ (cf. Eq.~(\ref{eq:ht})) with parameter set $\boldsymbol{\lambda}$, FIM is defined as
\begin{equation}\label{eq_FIM}
{{\Gamma }^{ij}} \equiv \left( \frac{\partial h}{\partial {{\lambda }_{i}}},\frac{\partial h}{\partial {{\lambda }_{j}}} \right)  \,. 
\end{equation}
For a monochromatic signal of frequency $f$, the noise-weighted inner product
$\left(a, b\right) \simeq \frac{2}{S_{n}\left(f\right)} \int_{0}^{T_{\rm obs}} a(t) b(t) dt$ \cite{Shah2012},
where $S_{n}(f)$ is the power spectral density of the instrumental noise at frequency $f$, and $T_{\rm obs}$ is the observation time.
The optimal signal-to-noise ratio (SNR) for signal detection is defined as ${\rm SNR} \equiv (h,h)^{1/2}$.
The root-mean-square (RMS) error of parameter ${\lambda_i}$ is estimated as $\Delta {\lambda_i} = \sqrt {{\Sigma_{ii}}}$,
where the covariance matrix $\Sigma$ is the inverse of the FIM, i.e., $\Sigma  = {\Gamma^{-1}}$.
We extend the parameter set in \cite{Jaranowski1999PhRvD}, as our work considers the spin-orbit coupling for a triaxial non-aligned NS, which results in a parameter set
$\boldsymbol{\lambda}=(\ln{h_{10}},\ln{h_{20}},\ln{h_{30}},\alpha,\sin{\delta},\ln{P_{\rm b}},\cos{\iota},\theta_S,\psi_{\rm p},\ln{\Omega_{\rm r}},\ln{\Omega_{\rm p}})$,
in which the amplitudes for the different waveform components are defined as
\begin{equation}
h_{10}=\frac{{2G{b^2}{I_3}\epsilon \gamma }}{{{c^4}D}}, h_{20}=\frac{{64G{b^2}{I_3}\epsilon \kappa }}{{{c^4}D}}, h_{30}=\frac{4Gb^2{I_3}\epsilon \gamma^2}{c^4D} \,.
\end{equation}
Logarithms are taken for some parameters because the relative errors for them are more meaningful than the absolute errors. For example, $\Delta \ln{h_{10}}=\Delta h_{10}/h_{10}$ is simply the relative error in $h_{10}$. 
Since $\boldsymbol{J}=\boldsymbol{L}+\boldsymbol{S}$, then $\theta_L$ is related to $\theta_S$ by $S/\sin(\theta_L)=L/\sin(\theta_S)$. According to Eq.~(\ref{eq_Omegapre}) and Kepler's third law, $\Omega_{\rm pre} \propto P_{\rm b}^{-5/3}$ and $r_1 \propto P_{\rm b}^{2/3}$. Therefore, the parameters $(\theta_L, \Omega_{\rm pre}, r_1)$ are related to $\theta_S$ and $P_{\rm b}$ used in the FIM, so they are not included in the parameter set $\boldsymbol{\lambda}$. 
Similar to the sky localization error defined in \citep{cutler1998lisa}, we define the corresponding one for a source located at $(\alpha,\delta)$, 
$ \Delta \Omega = 2 \pi \sqrt {\Sigma_{\alpha \alpha} \Sigma_{\sin \delta \sin \delta } -\Sigma_{\alpha \sin \delta }^{2} } $. 

In the following analysis, since the sensitivity of Einstein Telescope \cite{ET2010} in the frequency band of interest is not as good as that of Cosmic Explorer, we use Cosmic Explorer, which consists of two facilities (one 40 km on a side and one 20 km on a side), each with a single L-shaped detector \cite{CosmicExplorer2022}. 
Various angular parameters are taken as $\zeta=\pi/2, \lambda=0.764, \gamma_{\rm o}=1.5, \phi_{\rm r}=\phi_{\rm o}=0, \alpha=1.209, \delta=1.475, \psi_{\rm p}=1.0$.
Assuming a 40 km arm length with a low-frequency optimized sensitivity is used, the amplitude spectral density of the instrumental noise $\sqrt{S_{n}(f_0)} = 1.36\times 10^{-25}~{\rm Hz}^{-1/2}$ and $\sqrt{S_{n}(2f_0)} = 1.66\times 10^{-25}~{\rm Hz}^{-1/2}$ for $f_0=100~{\rm Hz}$, the observation time $T_{\rm obs}=2T_{\rm pre}=1.676 \times 10^7~{\rm s}$.

\begin{figure}[!htbp]
\centering
\includegraphics[scale=0.39]{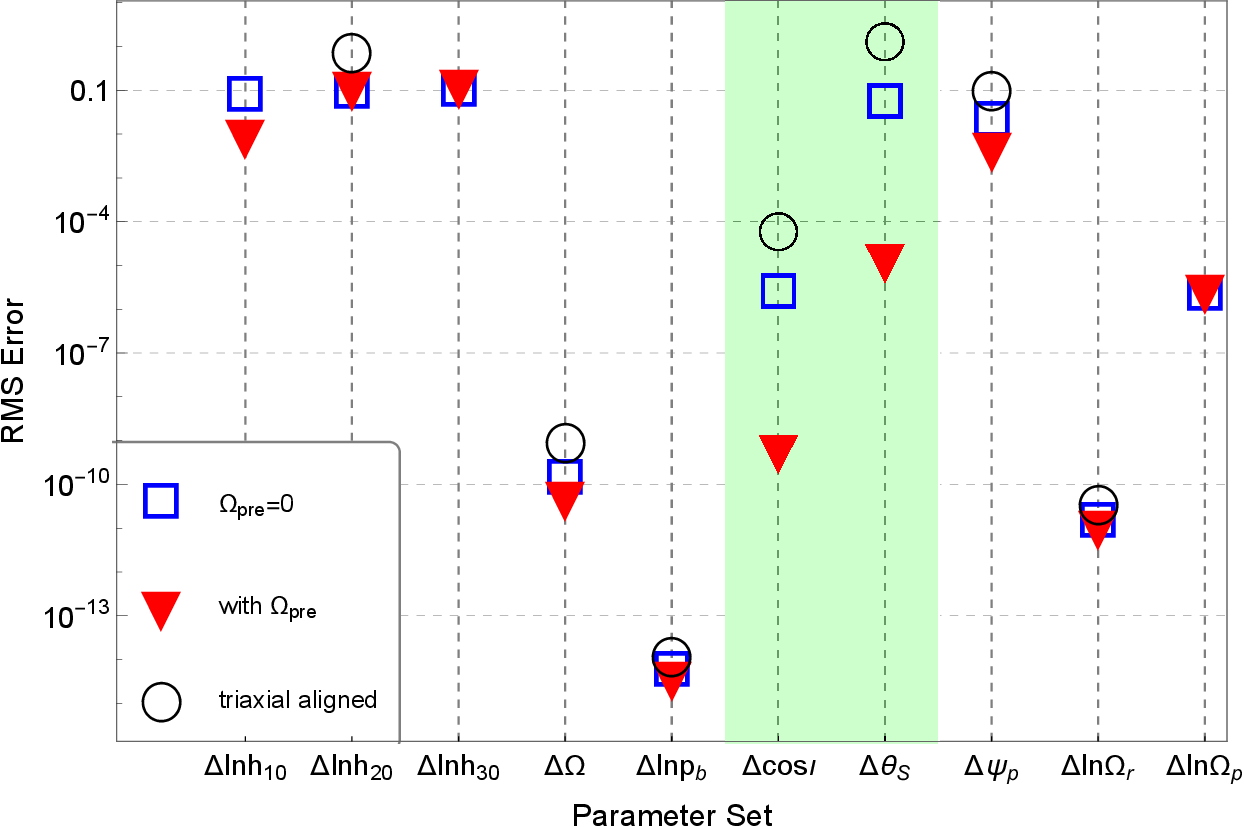}
\caption{The RMS errors (red downward-pointing triangles) of the signal parameters for a typical spinning NS in a binary using Cosmic Explorer detector. The effects of spin-orbit coupling are considered in the waveforms. For comparison, the blue open squares represent the result without spin-orbit coupling and the black circles represent the triaxial aligned case (the parameters $(h_{10},h_{30},\Omega_{\rm p})$ do not exist in this simplest model, cf. Sec.~\ref{sec:comparisonwithisolated}). The error of the sky location $(\alpha,\delta)$ is combined into the elliptical area $\Delta \Omega$. 
The parameter estimations for the triaxial non-aligned cases are more accurate than those for the triaxial aligned case. 
After taking into account spin-orbit coupling, the estimation accuracy of the parameters $\cos{\iota}$ and $\theta_S$ (in green shadow) are improved by about 3 orders of magnitude.}
\label{fig:RMS}
\end{figure}

After calculating the FIM numerically with Mathematica for Eq.~(\ref{eq_FIM}), the RMS errors of the estimated parameters for a typical spinning NS in a binary (the parameters used here are the same as in Fig.~\ref{fig:deltaOmegai} and Fig.~\ref{fig:h_inbinary_isolated}) are shown in Fig.~\ref{fig:RMS} as red downward-pointing triangles. For comparison, the results without spin-orbit coupling are represented by blue open squares and the results for the triaxial aligned case are shown as black circles.
For the first three Doppler-modulated signal components $h_n(t)~(n=1,2,3)$ in Eq.~(\ref{eq:ht}), their SNRs are 104, 8.5, and 7.6, respectively.  In comparison, the corresponding SNRs are 74.3, 12.6, and 11.3 for the signals without spin-orbit coupling. 
The total SNR for the triaxial aligned case is also 12.6 as that for $h_2$ without spin-orbit coupling, since the SNRs for $h_1$ and $h_3$ become zero when $\gamma=0$ (cf. Sec.~\ref{sec:comparisonwithisolated}). 
The spin and orbital precession modulated signal $h_1(t)$ increases its SNR by 40\% compared with the case without spin-orbit coupling, while the other two signals $h_2(t)$ and $h_3(t)$ both decrease by 33\%.
We can get some clues to understand the changes in SNRs from the limiting case of the waveforms in Sec.~\ref{sec:comparisonwithisolated}. From Eqs.~(\ref{eq_isoNS}) ($\iota$ corresponds to $\iota-\theta_S$ under spin-orbit coupling), we can see that $h_{+,\times}^{(1)}$ is proportional to $\sin(2\iota)$ or $\sin{\iota}$, while $h_{+,\times}^{(2)}$ and $h_{+,\times}^{(3)}$ are proportional to $(1+\cos^2{\iota})$ or $\cos{\iota}$. 
The spin-orbit coupling makes $\iota$ larger during one precession period (cf. $\iota-\theta_S$ in Fig.~\ref{fig:JS_coord}), therefore the SNR of $h_1(t)$ becomes larger and that of $h_2(t)$ and $h_3(t)$ becomes smaller.

The fractional estimation errors for three amplitudes are inversely proportional to their SNRs. 
So the parameter estimations for the triaxial non-aligned cases are more accurate than those for the triaxial aligned case. For the triaxial non-aligned case, the precession improves the sky localization by a factor of two and slightly improves the estimation of the orbital period. 
The improvement in sky localization is small because for double NS systems, whose orbital angular momentum is very close to the total angular momentum, the modulation of the orbital angular momentum is relatively small. 
The most significant improvement comes from the estimation of the angles $\cos{\iota}$ and $\theta_{S}$, both of which are improved by about 3 orders of magnitude. 
This is because these two angles are encoded in the amplitudes of the waveforms (cf. Eqs.~(\ref{eq:hinbinary})), and they modulate the profiles of the waveforms in Fig.~\ref{fig:h_inbinary_isolated}.

\section{Conclusions}\label{sec:conclusion}

In this work, we calculate the gravitational waveforms of a triaxial non-aligned NS in a compact binary system in which the effects of spin-orbit coupling have been incorporated. Then, we compare our waveforms with the ones commonly used in current CWs searches. Finally, we evaluated the parameter estimation accuracy for the signal detected by the proposed next-generation GW detector using the Fisher information matrix method.

For a tight double NS system with a 6-min orbital period, by solving the precession equation with the perturbation method, we find that spin precession-induced correction to the spin angular frequencies of NS is in the same order of magnitude as the angular frequency of orbital precession. The fitting factor between the waveforms with and without spin precession will drop to less than 0.97 after a few days. The analytic waveforms show that spin-orbit coupling introduces additional modulation information that will help in improving the accuracy of parameter estimation in CW detection.

The double NS system (consisting of a rapidly spinning NS and a nonspinning NS) considered in this work can be seen as a dual-line GW source, in which the orbital motion of the binary will emit low-frequency GWs in the mHz band in addition to the high-frequency GWs from the spinning NS. This dual-line GW source is of astrophysical interest, such as constraining the NS's moment of inertia and ellipticity using the ratio of the strain amplitudes of the low- and high-frequency GWs  \cite{Tauris2018} or combining the angular momentum loss of the NS \cite{Chen2021}. Since the angular frequency of the orbital precession contains information about the orbital period and mass of the binary, we can use it to infer binary parameters by combining information from the emitted GWs of the dual-line sources, and such studies are currently under our investigation.

\begin{acknowledgments}
Y.W. gratefully acknowledges support from the National Key Research and Development Program of China (No. 2022YFC2205201 and No. 2020YFC2201400), 
 the National Natural Science Foundation of China (NSFC) under Grants No. 11973024, Major Science and Technology Program of Xinjiang Uygur Autonomous Region (No. 2022A03013-4), and Guangdong Major Project of Basic and Applied Basic Research (Grant No. 2019B030302001). 
T. L. is supported by NSFC Grant No. 12003008 and the China Postdoctoral Science Foundation Grant No. 2020M682393.
J.-W. C. acknowledges the support from China Postdoctoral Science Foundation under Grant No. 2021M691146. S.D.M is supported by U.S. National Science Foundation (NSF) grant PHY-2207935. 
We thank the anonymous referee for helpful comments and suggestions. 

\end{acknowledgments}

\appendix

\begin{widetext}

\section{Expressions for \texorpdfstring{$\mathcal{R}$}{} and \texorpdfstring{$\mathcal{A}$}{}}
\label{App:eq:mathcal_R_and_A}

The symmetric matrix $\mathcal A$ used in calculating the waveforms (cf. Eqs. (\ref{eq:hinbinary})) can be explicitly expressed as follows 
%
\begin{subequations}
\label{eq:Amatrixinbinary}
\begin{align} 
\label{eq:mA11}
\mathcal{A}_{11} &=  2 b I_3 \epsilon  [16 b \kappa +\gamma \sin (t \Omega_{\rm p})[b \gamma  \sin (t \Omega_{\rm p}) +2 \Omega_{\rm pre} \sin \theta_S \sin (t \Omega_{\rm r})]] \,, \\
\mathcal{A}_{22} &=  2 b I_3 \epsilon  [-16 b \kappa +\gamma  \cos (t \Omega_{\rm p})[b \gamma  \cos (t \Omega_{\rm p}) +4 \Omega_{\rm pre} \sin \theta_S \sin^2 (t \Omega_{\rm r}/2)]] \,, \\
\mathcal{A}_{33} &=  -2 b \gamma I_3 \epsilon  [b \gamma +4 \Omega_{\rm pre} \sin \theta_S \sin (t \Omega_{\rm r}/2)\sin(t (2\Omega_{\rm p}+\Omega_{\rm r})/2)] \,, \\
\mathcal{A}_{12} &= -b \gamma I_3 \epsilon  [b \gamma  \sin (2 t \Omega_{\rm p})+2 \Omega_{\rm pre} \sin \theta_S [\sin (t \Omega_{\rm p})-\sin (t (\Omega_{\rm p}-\Omega_{\rm r}))]] \,, \\
\mathcal{A}_{23} &= b^2 I_3 \epsilon \gamma \sin (t \Omega_{\rm p}) \,, \\ \label{eq:mA31}
\mathcal{A}_{31} &= b I_3 \epsilon [b \gamma \cos (t \Omega_{\rm p}) +\Omega_{\rm pre} \sin \theta_S] \,. 
\end{align}
\end{subequations}
%

The transformation matrix $\mathcal R$ used in calculating the waveforms (cf. Eqs. (\ref{eq:hinbinary})) can be explicitly expressed as follows 
%
\begin{subequations}
\label{eq:Rinbinary}
\begin{align} \label{eq:Rx1}
{\mathcal{R}_{x1}} &= \sin (t{\Omega _{\rm{p}}})\Big[\cos (t{\Omega _{{\rm{pre}}}})\cos (t({\Omega _{\rm{p}}} + {\Omega _{\rm{r}}})) - \cos {\theta _S}\sin (t{\Omega _{{\rm{pre}}}})\sin (t({\Omega _{\rm{p}}} + {\Omega _{\rm{r}}}))\Big]+ \frac{1}{2}\cos (t{\Omega _{\rm{p}}}) \\ \nonumber
&\times \Big[\cos(t{\Omega _{{\rm{pre}}}})[16\kappa\sin(t{\Omega _{\rm{p}}}) \big(\cos(t{\Omega _{\rm{r}}}) + \kappa [3\cos (t(2{\Omega _{\rm{p}}} - {\Omega _{\rm{r}}}))- 8\cos (t{\Omega _{\rm{r}}}) + \cos (t(2{\Omega _{\rm{p}}} + {\Omega _{\rm{r}}}))]\big) \\ \nonumber
&+ ({\gamma ^2} - 2)\sin (t({\Omega _{\rm{p}}} + {\Omega _{\rm{r}}}))]+ \sin (t{\Omega _{{\rm{pre}}}})\big[ - 2\gamma \sin {\theta _S} + \cos {\theta _S}\Big(({\gamma ^2} - 2)\cos (t({\Omega _{\rm{p}}} + {\Omega _{\rm{r}}})) \\ \nonumber
&+ 16\kappa \sin (t{\Omega _{\rm{p}}})\big(3\kappa \sin (t(2{\Omega _{\rm{p}}} - {\Omega _{\rm{r}}})) + (8\kappa  - 1)\sin(t{\Omega _r})- \kappa \sin (t(2{\Omega _{\rm{p}}} + {\Omega _{\rm{r}}}))\big)\Big)\big]\Big] \,,\\
{\mathcal{R}_{x2}} &= \frac{1}{2}\sin (t{\Omega _{\rm{p}}})\Big[\sin (t{\Omega _{{\rm{pre}}}})(({\gamma ^2} - 2)\cos {\theta _S}\cos (t({\Omega _{\rm{p}}} + {\Omega _{\rm{r}}}))- 2\gamma (1 + 8\kappa )\sin {\theta _S}) \\ \nonumber
&+ ({\gamma ^2} - 2)\cos(t{\Omega _{{\rm{pre}}}})\sin (t({\Omega _{\rm{p}}} + {\Omega _{\rm{r}}}))\Big]- \cos (t{\Omega _{\rm{p}}})\Big[\cos {\theta _S}\sin (t{\Omega _{{\rm{pre}}}})(8\kappa [\cos (t{\Omega _{\rm{r}}}) + \kappa (3\cos (t(2{\Omega _{\rm{p}}} - {\Omega _{\rm{r}}})) \\ \nonumber
& - 8\cos (t{\Omega _{\rm{r}}}) + \cos (t(2{\Omega _{\rm{p}}} + {\Omega _{\rm{r}}})))]\sin(t{\Omega _{\rm{p}}}) - \sin (t({\Omega _{\rm{p}}} + {\Omega _{\rm{r}}}))) + \cos (t{\Omega _{{\rm{pre}}}})\Big(\cos (t({\Omega _{\rm{p}}} + {\Omega _{\rm{r}}})) \\ \nonumber
& + 8\kappa \sin(t{\Omega _{\rm{p}}})[\sin(t{\Omega _{\rm{r}}})+ \kappa ( - 3\sin (t(2{\Omega _{\rm{p}}} - {\Omega _{\rm{r}}})) - 8\sin(t{\Omega _{\rm{r}}}) + \sin (t(2{\Omega _{\rm{p}}} + {\Omega _{\rm{r}}})))]\Big)\Big] \,,\\
{\mathcal{R}_{x3}} &= \gamma \cos {\theta _S}(1 + 4\kappa  - 4\kappa \cos (2t{\Omega _{\rm{p}}}))\cos (t({\Omega _{\rm{p}}} + {\Omega _{\rm{r}}}))\sin (t{\Omega _{{\rm{pre}}}}) + \frac{1}{2}({\gamma ^2} - 2)\sin{\theta _S}\sin (t{\Omega _{{\rm{pre}}}}) \\ \nonumber
& + \gamma \cos (t{\Omega _{{\rm{pre}}}})(1 + 8\kappa {\sin ^2}(t{\Omega _{\rm{p}}}))\sin(t({\Omega _{\rm{p}}} + {\Omega _{\rm{r}}})) \,,\\ 
{\mathcal{R}_{y1}} &= \frac{1}{2}\Big[2(\cos (t({\Omega _{\rm{p}}} + {\Omega _{\rm{r}}}))\sin(t{\Omega _{\rm{p}}}) + 4\kappa [\cos (t{\Omega _{\rm{r}}}) + \kappa (3\cos (t(2{\Omega _{\rm{p}}} - {\Omega _{\rm{r}}}))- 8\cos (t{\Omega _{\rm{r}}}) \\ \nonumber
&+ \cos (t(2{\Omega _{\rm{p}}} + {\Omega _{\rm{r}}})))]\sin(2t{\Omega _{\rm{p}}}))\sin (t{\Omega _{{\rm{pre}}}})+ \cos (t{\Omega _{\rm{p}}})(2\gamma \cos (t{\Omega _{{\rm{pre}}}})\sin{\theta _S} + ({\gamma ^2} - 2)\sin (t{\Omega _{{\rm{pre}}}})\sin (t({\Omega _{\rm{p}}} + {\Omega _{\rm{r}}}))) \\ \nonumber
& + \cos {\theta _S}\cos (t{\Omega _{\rm{pre}}})\Big( - ({\gamma ^2} - 2)\cos (t{\Omega _{\rm{p}}})\cos (t({\Omega _{\rm{p}}} + {\Omega _{\rm{r}}}))+ 2\sin (t{\Omega _{\rm{p}}})\sin (t({\Omega _{\rm{p}}} + {\Omega _{\rm{r}}})) + 8\kappa \sin(2t{\Omega _{\rm{p}}})[\sin(t{\Omega _{\rm{r}}}) \\ \nonumber
& + \kappa ( - 3\sin (t(2{\Omega _{\rm{p}}} - {\Omega _{\rm{r}}})) - 8\sin(t{\Omega _{\rm{r}}}) + \sin (t(2{\Omega _{\rm{p}}} + {\Omega _{\rm{r}}})))]\Big)\Big] \,,\\ 
{\mathcal{R}_{y2}} &= \frac{1}{2}\Big[2\gamma (1 + 8\kappa )\cos(t{\Omega _{{\rm{pre}}}})\sin {\theta _S}\sin (t{\Omega _{\rm{p}}}) + \cos {\theta _S}\cos (t{\Omega _{{\rm{pre}}}})( - ({\gamma ^2} - 2)\sin (t{\Omega _{\rm{p}}})\cos (t({\Omega _{\rm{p}}} + {\Omega _{\rm{r}}})) \\ \nonumber
&+ 8\kappa [\cos (t{\Omega _{\rm{r}}}) + \kappa (3\cos (t(2{\Omega _{\rm{p}}} - {\Omega _{\rm{r}}})) - 8\cos (t{\Omega _{\rm{r}}}) + \cos (t(2{\Omega _{\rm{p}}} + {\Omega _{\rm{r}}})))]\sin (2t{\Omega _{\rm{p}}}) - 2\cos (t{\Omega _{\rm{p}}})\sin (t({\Omega _{\rm{p}}} + {\Omega _{\rm{r}}}))) \\ \nonumber
& + \sin (t{\Omega _{{\rm{pre}}}})\Big( - 2\cos(t{\Omega _{\rm{p}}})\cos(t({\Omega _{\rm{p}}} + {\Omega _{\rm{r}}})) + ({\gamma ^2} - 2)\sin (t{\Omega _{\rm{p}}})\sin (t({\Omega _{\rm{p}}} + {\Omega _{\rm{r}}}))+ 8\kappa \sin (2t{\Omega _{\rm{p}}})[3\kappa \sin (t(2{\Omega _{\rm{p}}} - {\Omega _{\rm{r}}})) \\ \nonumber
& + (8\kappa  - 1)\sin (t{\Omega _{\rm{r}}}) - \kappa \sin (t(2{\Omega _{\rm{p}}} + {\Omega _{\rm{r}}}))]\Big)\Big] \,,\\
{\mathcal{R}_{y3}} &= \gamma \cos {\theta _S}( - 1 - 4\kappa  + 4\kappa \cos (2t{\Omega _{\rm{p}}}))\cos (t({\Omega _{\rm{p}}} + {\Omega _{\rm{r}}}))\cos (t{\Omega _{{\rm{pre}}}}) - \frac{1}{2}({\gamma ^2} - 2)\sin{\theta _S}\cos (t{\Omega _{{\rm{pre}}}}) \\ \nonumber
&+ \gamma \sin (t{\Omega _{{\rm{pre}}}})(1 + 8\kappa {\sin ^2}(t{\Omega _{\rm{p}}}))\sin(t({\Omega _{\rm{p}}} + {\Omega _{\rm{r}}})) \,,\\ 
{\mathcal{R}_{z1}} &= \gamma \cos {\theta _S}\cos (t{\Omega _{\rm{p}}}) + \frac{1}{2}\sin{\theta _S}[({\gamma ^2} - 2)\cos (t({\Omega _{\rm{p}}} + {\Omega _{\rm{r}}}))\cos (t{\Omega _{\rm{p}}}) - 2\sin(t{\Omega _{\rm{p}}})\sin(t({\Omega _{\rm{p}}} + {\Omega _{\rm{r}}})) \\ \nonumber
& + 8\kappa \sin (2t{\Omega _{\rm{p}}})(3\kappa \sin(t(2{\Omega _{\rm{p}}} - {\Omega _{\rm{r}}}))+ (8\kappa  - 1)\sin(t{\Omega _{\rm{r}}}) - \kappa \sin(t(2{\Omega _{\rm{p}}} + {\Omega _{\rm{r}}})))] \,,\\
{\mathcal{R}_{z2}} &= \gamma (1 + 8\kappa )\cos {\theta _S}\sin (t{\Omega _{\rm{p}}}) + \frac{1}{2}\sin{\theta _S}\Big[({\gamma ^2} - 2)\cos (t({\Omega _{\rm{p}}} + {\Omega _{\rm{r}}}))\sin (t{\Omega _{\rm{p}}})+ 2\cos (t{\Omega _{\rm{p}}})\sin(t({\Omega _{\rm{p}}} + {\Omega _{\rm{r}}})) \\ \nonumber
& - 8\kappa [\cos (t{\Omega _{\rm{r}}}) + \kappa (3\cos (t(2{\Omega _{\rm{p}}} - {\Omega _{\rm{r}}})) - 8\cos (t{\Omega _{\rm{r}}}) + \cos (t(2{\Omega _{\rm{p}}} + {\Omega _{\rm{r}}})))]\sin(2t{\Omega _{\rm{p}}}) \Big] \,,\\ \label{eq:Rz3}
{\mathcal{R}_{z3}} &= \frac{1}{2}(2 - {\gamma ^2})\cos {\theta _S} + \gamma \cos (t({\Omega _{\rm{p}}} + {\Omega _{\rm{r}}}))\sin {\theta _S}(1 + 8\kappa \sin^2(t{\Omega _{\rm{p}}})) \,.  
\end{align}
\end{subequations}

\section{Expressions for \texorpdfstring{$O(\Omega_{\rm pre})$}{} waveform components}
\label{App:eq:h+xpre}

The waveform components to the order $O(\Omega_{\rm pre})$ can be expressed as follows 
%
\begin{subequations}
\label{eq:hpre}
\begin{align} \label{eq:h+pre}
h_ + ^{({\rm pre})} &= \frac{{Gb{I_3}\epsilon {\Omega _{{\rm{pre}}}}\sin {\theta _S}}}{{4{c^4}D}}[\cos(t\Omega _{\rm{r}})(\sin (2{\theta _S})(-(3 + \cos (2\iota ))\cos(2t{\Omega _{{\rm{pre}}}})+6 {\sin ^2}\iota) \\ \nonumber
 &+ 4\cos (2{\theta _S})\cos (t{\Omega _{\rm{pre}}})\sin(2\iota)) + 2(-2\cos{\theta _S}\sin(2\iota )\sin(t{\Omega _{{\rm{pre}}}}) + (3 + \cos (2\iota))\sin {\theta _S}\sin (2t{\Omega _{{\rm{pre}}}}))\sin (t{\Omega _{{\rm{r}}}})] \,, \\ \label{eq:hxpre}
h_{\times} ^{({\rm{pre}})} &= \frac{{Gb{I_3}\epsilon {\Omega _{{\rm{pre}}}}\sin {\theta _S}}}{{{c^4}D}}[\cos (t\Omega _{\rm{r}})(2\cos (2{\theta _S})\sin \iota \sin (t{\Omega _{{\rm{pre}}}})- \cos \iota \sin (2{\theta _S})\sin (2t{\Omega _{{\rm{pre}}}}))  \\ \nonumber 
 &+ 2( - \cos \iota \sin {\theta _S}\cos (2t{\Omega _{{\rm{pre}}}}) + \cos {\theta _S}\cos (t{\Omega _{{\rm{pre}}}})\sin \iota )\sin (t\Omega _{\rm{r}})] \,.
\end{align}
\end{subequations}
%
We can see that there are  components with frequencies of $\Omega_{\rm r}$, $\Omega_{\rm r} \pm \Omega_{\rm pre}$, and $\Omega_{\rm r} \pm 2 \Omega_{\rm pre}$ in these waveforms. 
Since ${h_ {+,\times} ^{({\rm pre})}}/{h_ {+,\times} ^{(1)}} \sim {\Omega_{\rm pre}\sin\theta_S}/{(b \gamma)}, ~{h_ {+,\times} ^{({\rm pre})}}/{h_ {+,\times} ^{(2)}} \sim {\Omega_{\rm pre}\sin\theta_S}/{(16 b \kappa)}$, and ${h_ {+,\times} ^{({\rm pre})}}/{h_ {+,\times} ^{(3)}} \sim {\Omega_{\rm pre}\sin\theta_S}/{(b \gamma^2)}$, for typical parameters used as in Fig. \ref{fig:deltaOmegai},  these $h_ {+,\times} ^{({\rm pre})}$ components are  negligible compared to $h_{+,\times}^{(1)}$,  $h_{+,\times}^{(2)}$, and $h_{+,\times}^{(3)}$.

\end{widetext}

\section{The residual of two solutions}
\label{App:residual}

\begin{figure}[!htbp]
\centering
\includegraphics[scale=0.6]{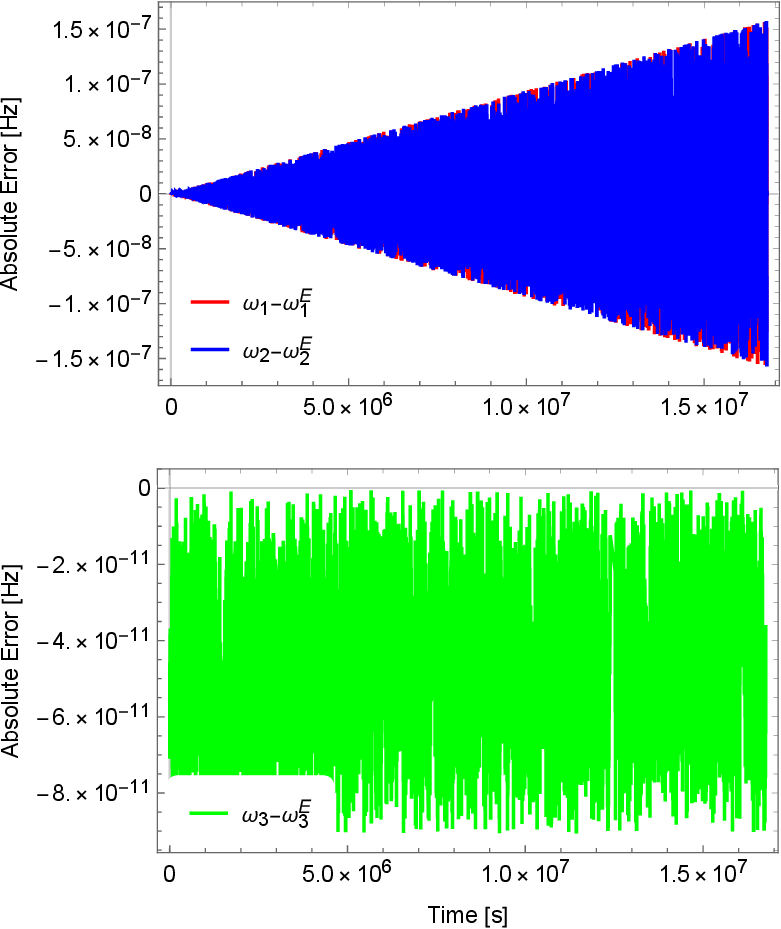}
\caption{The absolute errors between the approximate angular frequency $\omega_i$ and the exact angular frequency $\omega_i^E$ during two orbital precession periods. The parameters used here are the same as in Fig. \ref{fig:deltaOmegai}.}
\label{fig:residual}
\end{figure}

To confirm the fidelity of the approximate calculation, we use Euler angles to accurately calculate the angular frequency. Although this method can give the exact solution $\omega_i^E$, it is too complicated to give a simple analytical waveform like Eqs.~(\ref{eq:hinbinary}). They can be calculated as (an overdot represents $d/dt$) \cite{LANDAU1976}
\begin{subequations}
\begin{align}
\omega_{1}^E &= \dot{\phi}_{\rm b} \sin \theta_{\rm b} \sin \psi_{\rm b} + \dot{\theta}_{\rm b} \cos \psi_{\rm b} \,, \\
\omega_{2}^E &= \dot{\phi}_{\rm b} \sin \theta_{\rm b} \cos \psi_{\rm b}-\dot{\theta}_{\rm b} \sin \psi_{\rm b} \,, \\
\omega_{3}^E &= \dot{\phi}_{\rm b} \cos \theta_{\rm b} + \dot{\psi}_{\rm b} \,,
\end{align}
\end{subequations}
with the Euler angles derived from the primitive (no approximation) rotation matrix (cf. Eq.~(\ref{eq_newR})), 
\begin{subequations}
\begin{align}
\theta_{\rm b} &= \arccos \left(\mathcal{R}_{z3} \right) \,, \\
\phi_{\rm b}   &= \arctan \left(-\frac{\mathcal{R}_{x3}}{\mathcal{R}_{y3}}\right) \,, \\
\psi_{\rm b}   &= \arctan \left(\frac{\mathcal{R}_{z1}}{\mathcal{R}_{z2}}\right) \,.
\end{align}
\end{subequations}
The absolute errors between the approximate angular frequency $\omega_i$ and the exact angular frequency $\omega_i^E$ are shown in Fig.~\ref{fig:residual}. 
During two orbital precession periods, the relative deviation of the analytic approximation from the exact solution is $\lesssim 10^{-9}$ for $\omega_{1,2}$ and $\lesssim 10^{-13}$ for $\omega_{3}$. Therefore, the solution of the angular frequency is accurate enough for the calculation of the waveforms.

\section{Effects of orbital precession on Doppler shift}
\label{sec:Dop_shift_cor}

\begin{figure}[!htbp]
\centering
\includegraphics[scale=0.21]{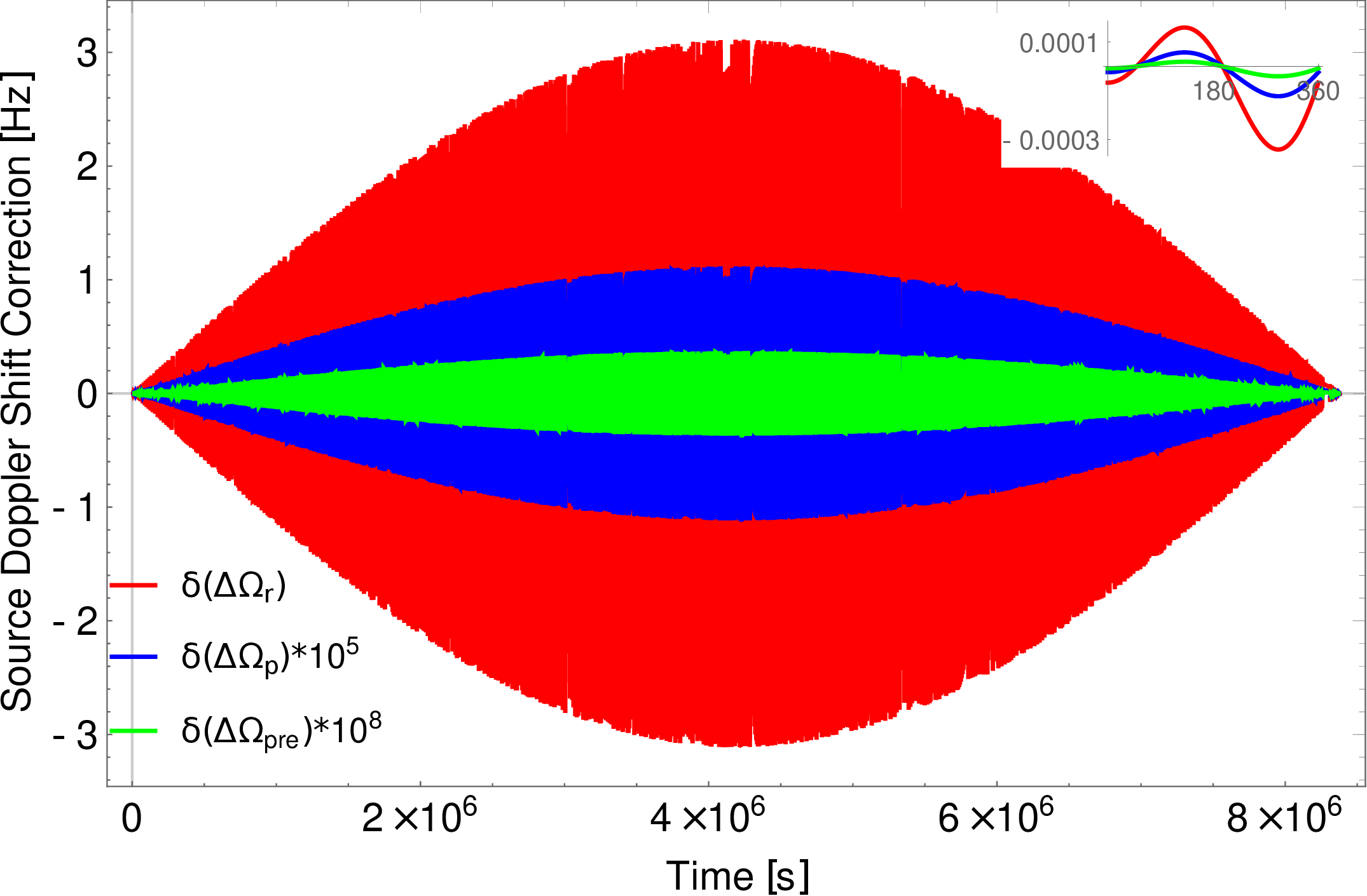}
\caption{The Doppler shift corrections to the different frequency components in the waveforms caused by orbital plane precession during one orbital precession period. For visualisation purposes, $\delta(\Delta\Omega_{\rm p})$ and $\delta(\Delta\Omega_{\rm pre})$ are magnified by a factor of $10^5$ and $10^8$, respectively. The parameters used here are the same as in Fig. \ref{fig:deltaOmegai}.} 
\label{fig:Dop_Correction}
\end{figure}

The Doppler shift correction is defined as 
\begin{equation}
 \delta (\Delta \Omega_{n}) = \Delta \Omega_{n} - \Delta \Omega_{n}(\Omega_{\rm pre}=0) 
\end{equation}
with $n=\{\rm r, p, pre\}$. It measures the effects of spin-orbit coupling on the phase of GWs of the spinning NS in a binary. 
As seen in Fig.~\ref{fig:Dop_Correction}, the deviation of the GW frequency can reach $\sim 0.5\%$ for $f_0$ and $\sim 1\%$ for $2f_0$ if we do not consider the orbital plane precession.

%

\end{document}